\documentclass[11pt]{scrartcl}
\pdfoutput=1
\usepackage[english]{babel}%
\usepackage{amsmath,amssymb,amsfonts,graphicx,multicol} 
\usepackage{mathrsfs}
\usepackage[numbers,sort&compress]{natbib}
\usepackage[utf8]{inputenc}%
\usepackage[font=small,labelfont=bf,format=plain, margin=0em, indention=0em]{caption} %
\usepackage[usenames, dvipsnames]{xcolor} %
\usepackage[citecolor=OliveGreen, linkcolor=blue, urlcolor=OliveGreen, colorlinks=true]{hyperref}
\usepackage{hypcap}
\usepackage[font=footnotesize]{subcaption}
\usepackage{pdflscape}
\usepackage[shortlabels]{enumitem}
\usepackage{multirow}
\usepackage{hhline}
\usepackage{hyperref}

\usepackage{float}
\restylefloat{table}

\definecolor{nicered}{rgb}{0.7,0.1,0.1}
\definecolor{nicegreen}{rgb}{0.1,0.5,0.1}
\hypersetup{colorlinks,citecolor= nicegreen,linkcolor= nicered}

\setkomafont{sectioning}{\normalfont\bfseries}

\newcommand{\ifb}{\text{fb}^{-1}}
\newcommand{\tn}{\textnormal}

\newcommand{\GeV}{\ensuremath{\,\text{GeV}}}
\newcommand{\TeV}{\ensuremath{\,\text{TeV}}}
\newcommand{\gr}[1]{\textcolor{gray}{}}
\newcommand{\mc}[1]{\multicolumn{2}{|c|}{#1}}

\begin{document}
\begin{titlepage}
\vspace*{-2.cm}
\begin{flushright}
{\small
TUM-HEP 1054/16
}
\end{flushright}
\vspace*{1.cm}

\begin{center}
{\Large \textbf{
On the challenge of estimating diphoton backgrounds at large invariant mass
}}
\end{center}
\vskip0.5cm

\renewcommand{\thefootnote}{\fnsymbol{footnote}}

\begin{center}
{\large J.F.~Kamenik$^{a,b}$, G.~Perez$^c$, M.~Schlaffer$^c$, A.~Weiler$^d$}
\end{center}

\vskip 20pt

\def\LjubljanaFMF{Faculty of Mathematics and Physics, University of Ljubljana,\\ Jadranska 19, 1000
Ljubljana, Slovenia }
\def\LjubljanaIJS{Jo\v zef Stefan Institute, Jamova 39, 1000 Ljubljana, Slovenia}
\def\CERN{CERN, Theory Division, CH-1211 Geneva 23, Switzerland}
\def\Weizmann{Department of Particle Physics and Astrophysics,
Weizmann Institute of Science, \\7610001 Rehovot, Israel}
\def\MunichTU{Physik Department T75, Technische Universit\"at M\"unchen, James-Franck-Strasse 1,  \\85748 Garching, Germany}

\begin{center}
  $^a$ \LjubljanaIJS \\
  $^b$ \LjubljanaFMF \\
  $^c$ \Weizmann\\
  $^d$ \MunichTU\\
\end{center}

\vglue .5truecm

\begin{abstract}
  \footnotesize
  \noindent Diphoton searches at high invariant mass are an integral part of the experimental high
  energy frontier. Using the analyses of the 750\,GeV diphoton resonance as a case study, we examine
  the methodology currently employed by the experimental analyses in estimating the dominant
  standard model backgrounds. We assess the dependence of the significance associated with the
  excess on the background modeling. In particular, we show that close to the high energy tails of
  the distributions, estimates of the jet faking backgrounds relying on functional extrapolations or
  Monte Carlo estimates of the challenging photon-jet contributions introduce a large
  uncertainty. Analyses with loose photon cuts, low photon $p_T$ cuts and those susceptible to high
  photon rapidity regions are especially affected. Given that diphoton-based searches beyond 1\,TeV
  are highly motivated as discovery modes, these considerations are relevant and applicable in the
  future.\par
  We first consider a physics-driven deformation of the shape of the diphoton faking photon-jet
  spectrum by next-to-leading order effects combined with a rapidity and transverse momentum
  dependent fake rate. We show that the resulting local significance of the excess is reduced due to
  such a deformation. Then using a simple, but more general, ansatz to modify the contribution of
  the jet faking backgrounds at high invariant masses but keeping the inclusive and differential
  sample purities within experimental uncertainty estimates, we demonstrate that the originally
  reported local 750\,GeV excess significances could have been overestimated by more than one
  standard deviation. We furthermore cross-check our analysis by comparing fit results based on the
  smaller 2015 and the larger 2016 LHC datasets. Finally we employ our methodology on the complete
  13\,TeV LHC dataset to asses the systematics involved in the current diphoton searches beyond the
  TeV region.
\end{abstract}

\end{titlepage}

\section{Introduction}
\label{sec:introduction}

Searches for new physics at the energy frontier often look for new phenomena at the edge of
distributions. In this kinematical region the knowledge of the Standard Model (SM) background is
typically limited and the challenge is to look for a new resonance where only partial knowledge on
the SM background is available. In this paper we focus in particular on new physics probes based on
the high diphoton invariant mass spectrum. We examine, using the analyses of the 750\,GeV diphoton
resonance as a case study, the strategy currently used by the experimental collaborations in
estimating the dominant SM backgrounds. We employ our methodology on the 13\,TeV LHC dataset to
asses the systematics involved in the current diphoton searches beyond the TeV region\par

In their 2015 data sets, both ATLAS and CMS observed an excess in the diphoton spectrum near
$m_{\gamma \gamma}=750\,\GeV\,.$ The relevant details of the ATLAS and CMS analyses are described
in~\cite{Aaboud:2016tru, Khachatryan:2016hje}. At face value the local significances for a broad
resonance were given by
\begin{equation}
  \label{sigs}
  \begin{aligned}[c]
    p_{\textrm{ATLAS}}&=4 \times 10^{-5}\,,\\
    \sigma_{\textrm{ATLAS}}& = 3.9\,,
  \end{aligned}
  \qquad
  \begin{aligned}[c]
    p_{\textrm{CMS}} &= 5\times 10^{-3}\,,\\
    \sigma_{\textrm{CMS}} &= 2.6\,,
  \end{aligned}
  \qquad
  \begin{aligned}[c]
    p_{\textrm{comb}}&= 1 \times 10^{-6}\,,\\
    \sigma_{\textrm{comb}}&= 4.7\,,
  \end{aligned}
\end{equation}
where $p_{\textrm{ATLAS, CMS, comb}}$ $(\sigma_{\textrm{ATLAS, CMS, comb}})$ correspond to the local
$p$ value (confidence level) of ATLAS, CMS, and their naive combination.\footnote{We do not discuss
  here the global significance as it strongly depends on the lower value of $m_{\gamma \gamma}$
  defined for the search region. ATLAS (CMS) chose it to be about 200\,GeV (400\,GeV). Furthermore,
  as discussed below, the region below 500\,GeV is dominating the fit to the functional form which
  is used to estimate the background. Thus, it is not clear whether one should consider this region
  as a control region or as the region of interest for the search itself.}

The local results quoted in Eq.~\eqref{sigs} are quite significant and captured the attention of the
high energy community.  Interpreting them naively, one would be lead to one of the following
conclusions
\begin{enumerate}[(i)]
\item this excess is a result of a rare statistical fluctuation;
\item \label{NPclaim} this excess implies a discovery of non-Standard Model dynamics.
\end{enumerate}
As both conclusions are quite extraordinary (certainly the second one), they motivate an
investigation into their robustness.  In particular, we raise a third option, to be considered in
conjunction with (i), namely, we ask how unlikely is the possibility that
\begin{enumerate}[(i),resume]
\item \label{BKGclaim} the significance of the excess is overestimated due to underestimating
  fake-based backgrounds.
\end{enumerate}
With the inclusion of more data in the analyses the excess eventually vanished~\cite{ATLAS:2016eeo,
  CMS:2016crm}, ruling out the new physics hypothesis \ref{NPclaim}. However, the possibility of
claim \ref{BKGclaim} remains unclear, affecting all analyses which rely on a precise knowledge of
the photon faking background and use the same techniques to estimate it.\par
While our conclusion is independent of the 750\,GeV resonance we use it as an example case to
scrutinize the hypothesis of the underestimated background and its implications. First, the main
rationale behind our hypothesis is presented in Sec.~\ref{sec:rationale}, followed by a detailed
description of our approach to background estimation (Sec.~\ref{sec:setup}) and the statistical
treatment of the data (Sec.~\ref{sec:stat-treatm}). The comparison with the full 2016 data set is
presented in Sec.~\ref{sec:post-ichep-update}. Our main conclusions are summarized in
Sec.~\ref{sec:conclusions}. For other relevant works, see
Refs.~\cite{Davis:2016hlw,Bondarenko:2016rvd}.

\section{The rationale}
\label{sec:rationale}

Superficially, the experimental situation related to the diphoton excess was fairly
straightforward. The experiments had reported a relatively narrow ``bump'',
$\Gamma/m\lesssim 6\%\ll1$. Such a bump implies a rise in the differential distribution while, due
to the rapidly falling parton luminosity functions, it is expected that any reasonable
background-related distribution should be a monotonically decreasing function of the invariant
mass. Consequently, the presence of a non-Standard Model feature seemed to have been indicated by
the measurements. While this was qualitatively correct the challenge is to quantify the significance
of the excess. To endow the bump with a significance, one needs to control and quantify the
background.\par
The following approaches can be used to constrain the form of the background:\\

\textbf{I.} \textit{Data-driven approach.}  Assuming $\Gamma/m\ll1$ and a featureless monotonic
background, a robust way to constrain it is through \emph{interpolation} via a two-sided side band
analysis.  However, this requires to have enough measured events at invariant masses both below the
resonance and above it. In the case of the 750\,GeV excess, there were less than 40 events in all of
the analyses measured with invariant masses above 850 GeV. Such a small number of events does not
allow one to use this method reliably. \\

\textbf{II.} \textit{"First-principle"/Monte-Carlo approach.}  There is a rather narrow class of
observables for which the theory has reached an advanced enough level such that we can fully trust
our ability to correctly predict the shape of the background distributions.  We believe that the
invariant mass distribution of experimentally measured diphoton events does not (yet) belong to this
selected class of observables. Namely, the continuous diphoton distribution consist of an admixture
of two dominant components: (i) The first is made of two real isolated hard photons.  This diphoton
distribution is currently known to next-to-next-to-leading order (NNLO)
accuracy~\cite{Catani:2011qz,Campbell:2016yrh} in perturbative QCD and imposing cuts similar to the
ATLAS spin-0 analysis suggests an overall uncertainty of about 5\% for the invariant mass
distribution~\cite{Campbell:2016yrh}.  (ii) An additional important background component is due to
fakes coming mostly from processes involving a hard photon and a jet that passes the various photon
quality and isolation cuts~\cite{Aad:2016xcr}. In addition, depending on these cuts, also the dijet
background could play an important role. The prompt photon-jet cross section is currently known at
next-to-leading order (NLO) in QCD, and several codes are available to produce the relevant
distributions, including JetPhox~\cite{Catani:2002ny} and PeTeR~\cite{Becher:2013vva}. In addition,
QCD threshold resummation at next-to-next-to-next-to-leading logarithmic (N${}^3$LL)
order~\cite{Becher:2012xr,Becher:2015yea} as well as electroweak Sudakov effects are being
included~\cite{Schwartz:2016olw}, resulting in theory uncertainties of about
10-20\%~\cite{Schwartz:2016olw,Aad:2016xcr,Khachatryan:2015ira}. However, a comparison with the
8\,TeV ATLAS measurement~\cite{Aad:2016xcr} shows that at low photon $p_T\sim 50\,$GeV the data
exhibits some level of deviations from the theoretical predictions (a larger uncertainty is found
for the invariant mass distribution, see~\cite{Aad:2013gaa}).  In addition, it is important to note,
that the fake rate strongly depends on the quark/gluon ``flavor" of the tagged jet (for some
discussion on jet flavor definitions
see~\cite{Banfi:2006hf,Buckley:2015gua,Larkoski:2014pca,Bauer:2013bza}): intuitively one can
understand the difference through the quark and gluon fragmentation functions to pions. At large
$x$, as required to be able to pass photon isolation criteria, gluon fragmentation to few pions is
much more suppressed (see e.g. Chapter 20 in Ref.~\cite{Agashe:2014kda}). Accordingly, a dedicated
ATLAS study~\cite{ATLAS:2011kuc} found that there is a probability of about $1:2\times10^3$ for a
quark jet to fake a photon, and only $1:2\times10^4$ for a gluon jet to fake a photon, for jets with
$E_T>40\,$GeV.  Applying this to the photon-jet background, we also note that subleading jets might
become an important source of fakes if the leading jet is predominantly gluon-initiated.\par

In order to theoretically predict the purity of the diphoton mass distribution, an appropriate
admixture of the diphoton and the photon-jet(s) components needs to be
constructed~\cite{Neufeld:2010fj}.  Furthermore, for the latter component, one is required to
convolve the photon-jet distribution with the relevant fragmentation functions or at least tag the
flavor of the jet(s).  It is also important to note that the purity is a highly phase space
dependent quantity. Not only does it depend on the ratio of the differential jet-photon and
photon-photon production but also on the jet-to-photon fake rate. The fake rate may exhibit strong
dependence on the differential quantities such as $p_T$ and (pseudo)rapidity $\eta$.  For instance,
as discussed below, in the CMS analyses purity is estimated to be better than 90\% in the
(central-central) EBEB event category but only better than 80\% in the (forward-central) EBEE
one. Both experiments consider the purity in an inclusive way. However, in the relevant kinematical
region the data is not sufficient to constrain possibly large deviations from the inclusive purity
estimation (see Fig.~\ref{fig:purity}).\\

\textbf{III.}  \textit{Functional-fit approach.} Given the present practical limitations of the
methods \textbf{I} and \textbf{II}, one is lead to a more phenomenological approach in which the
background estimate is obtained by fitting an universal function to control regions in the data and
then \emph{extrapolating} into the signal regions using the fitted functional form.  This allows one
to predict the background at relatively high invariant masses in a straightforward
manner. Consequently, both experiments are essentially following this approach in most of their
analyses\footnote{An exception is the ATLAS spin-2 analysis which employs a Monte Carlo approach
  (\textbf{II}) with a data-driven estimate of the photon-jet and jet-jet background, see Section
  \ref{sec:stat-treatm}.}, although the functional forms used by ATLAS in the spin-0 analysis and by
CMS are slightly different.  Thus, the significance of the excess is mostly determined by comparing
measured events to a background estimate predicted by a fitting function.\\

While method \textbf{III} is very transparent and makes the search for bumps easy to analyze, it is
also rather susceptible to systematic effects, in particular a lack of understanding of the physics
modifying the tails of the distributions, as we argue below.  The fitting functions used by ATLAS
and CMS are well suited for describing rapidly falling distributions and are fitted to the available
data. With the amount of data in the 2015 data sets, the differentially measured number of events is
abundant in the low invariant mass region and is spare in the high mass region. The extraction of
the functions' parameters is thus dominantly controlled by the low $m_{\gamma \gamma}$ region and
hardly affected by modifications of the invariant mass distribution at diphoton masses of above
roughly 500\,GeV.  However, the significance of the excess with respect to the fitting function is
very much affected by such deformations. As it is hard to directly test or predict the correct form
of the diphoton mass distribution this raises the
following questions: \\
\textit{ ${\mathcal{I}}$.  Is the experimental signal over background estimation robust against the
  presence of deviations  from the fitting function predictions at large invariant masses?  \\
  ${\cal II}$. If this is not the case, can one produce smoking-gun predictions to show that indeed
  the significance of the excess is being overestimated?}

Let us first focus on point ${\cal I}$. To examine the sensitivity of the significance of the excess
to the variation of the tails of the distributions.  We consider a family of background shapes that
are formed by an admixture of the diphoton and photon-jet distributions. We keep the overall
inclusive purity of the samples at 90\% and 80\%, respectively, in accordance with the measured data
at low invariant masses.  More specifically, we use two classes of deformations.  The first is
derived from a modification of the photon-jet spectrum due to NLO and showering effects combined
with an increased fake rate for larger transverse momenta and pseudo-rapidities of the jets.\par

We then consider a simpler ansatz where we allow the distribution of the $p p \to \gamma j$
component to be reweighted at invariant masses above 500\,GeV such that the purity of events with
large invariant masses is reduced leading to a controlled deviation from the functional fit.  In the
following section we provide a detailed description of our approach. We also provide some tests of
our procedure to check that our method complies with public data (below and above the resonance
region) and is passing the relevant statistical tests.  We then report how the significance is
affected by the amount of rescaling of the distributions of fakes.  Finally we can use our ansatz to
address item ${\cal II}$ and provide smoking guns to test our hypothesis on overestimating the
excess significance. With the full statistics of the 2016 data sets at hand it would be fairly easy
to eliminate our hypothesis.

\section{Reducible and irreducible backgrounds}
\label{sec:setup}

The main background to the diphoton signal is the irreducible $ p p \to \gamma\gamma$ background. We
consider in the following: ATLAS spin-0 and spin-2, and CMS $13\,\TeV$ EBEB and EBEE categories with
magnets on. We generate the diphoton invariant mass spectrum at NNLO with MCFM version 8.0
\cite{Campbell:1999ah,Campbell:2011bn,Campbell:2015qma,Boughezal:2016wmq,Campbell:2016yrh} applying
the cuts as described in the respective analyses, see Table \ref{tab:cuts}. The main contribution to
the reducible background is the $p p \to \gamma j$ production where the hard jet is a quark jet that
is wrongly reconstructed as a photon. We generate this background at leading order (LO) with
MadGraph5 version 5.2~\cite{Alwall:2014hca}. We note that at LO the $p p \to \gamma j$ sample is
dominated by quark jets, which as already mentioned, lead to a much larger fake rate than gluon
jets.
\begin{table}[tb]
  \centering
  \begin{tabular}[t]{|l|c|c|c|c|}
    \hline
    Analysis             & ATLAS spin-0           & ATLAS spin-2 & CMS EBEB     & CMS EBEE  \\\hline
    $m_{\gamma\gamma}$     & $> 150\,\GeV$          & $>200\,\GeV$ & $>230\,\GeV$ & $>330\,\GeV$  \\
    $p_{T,1}$             & $>0.4\,m_{\gamma\gamma}$ & $>55\,\GeV$  & $>75\,\GeV$  & $>75\,\GeV$   \\
    $p_{T,2}$             & $>0.3\,m_{\gamma\gamma}$ & $>55\,\GeV$  & $>75\,\GeV$  & $>75\,\GeV$   \\
    $\left|\eta_1\right|$ & $<2.37$               & $<2.37$      & $<1.44$      & $<1.44$    \\
    $\left|\eta_2\right|$ & $<2.37$               & $<2.37$      & $<1.44$      & $1.57<\eta_2<2.5$ \\
    $\left|\eta\right|$ excluded & $1.37<\eta_{1,2}<1.52$ & $1.37<\eta_{1,2}<1.52$ & n.a. & n.a. \\\hline
    $\sigma_{\gamma\gamma}$ [pb] (NNLO) & 2.7             & 1.9         & 0.52           & 0.23\\
    $\sigma_{\gamma j}$ [pb] (NLO) & 1400          & 1000 & 250 &  130 \\
    $\left.{\sigma_{\gamma j}}{}/{\sigma_{\gamma\gamma}}{}\right|_{m>500\,\GeV_{}}$  & 510 &
                                                                                                     670
                                                                 & 470 & 640 
    \\\hline
  \end{tabular}
  \caption{Cuts of the analyses where the subscript refers to the hardest and second hardest photon
    candidate, the cross section of the $p p \to \gamma\gamma$ sample passing these cuts (calculated
    at NNLO with MCFM) and of the $p p \to \gamma j$ sample at hadron level (before applying any
    photon mistag rate), calculated at NLO with MadGraph5\_aMC@NLO, showered with
    Pythia~\cite{Sjostrand:2006za} and the jets clustered with anti-$k_T$, $R=0.4$ algorithm using
    FastJet~\cite{Cacciari:2011ma}. In the last line the ratios of the two distributions in the
    invariant mass region above 500\,GeV are given.}
  \label{tab:cuts}
\end{table}

The reconstructed diphoton distribution is a mixture of $p p \to \gamma\gamma$ and
$p p \to \gamma j$ invariant mass distributions.  Let us define a short-hand notation for the
normalized invariant mass distribution
\begin{equation}
  \label{eq:2}
  w_{\gamma X}\equiv \frac{1}{\sigma_{\gamma X}} \times \frac{\tn{d}\sigma_{\gamma X}}{{\tn{d} m_{\gamma X}}}\,,  
\end{equation}
with $X=\gamma,j$. The mixed distribution $w_\tn{mix}$ is a function of the normalisation
$N_\tn{mix}$ and a parameter $\mathcal{R}$, that controls the shape modification of $w_{\gamma j}$
and will be defined Eq.~\eqref{eq:1}. We write $w_\tn{mix}$ as
\begin{equation}
  \label{eq:5}
  w_\tn{mix}(N_{\textrm{mix}}) =N_{\textrm{mix}}  \left[\mathcal{P}w_{\gamma\gamma}+(1-\mathcal{P})w_{\gamma j} \right]\,,
\end{equation}
where $\mathcal{P}$ is the inclusive purity of the sample. We set $\mathcal{P}=90\%$ (80\%) for the
ATLAS and CMS EBEB (EBEE) analyses, which is within the reported error bands.  We will assume, that
$w_{\gamma\gamma}$ is obtained by normalizing the MCFM diphoton invariant mass distribution.  As for
$w_{\gamma j}$, following the rationale described in Section \ref{sec:rationale}, we modify in two
different ways as we now describe in detail.

\subsection{QCD and jet-fake dependence of the diphoton shape}
\begin{figure}[tb]
  \centering
  \begin{subfigure}[t]{0.48\linewidth}
    \begin{minipage}[t]{1.0\linewidth}
      \vspace{5pt}\includegraphics[width=\textwidth]{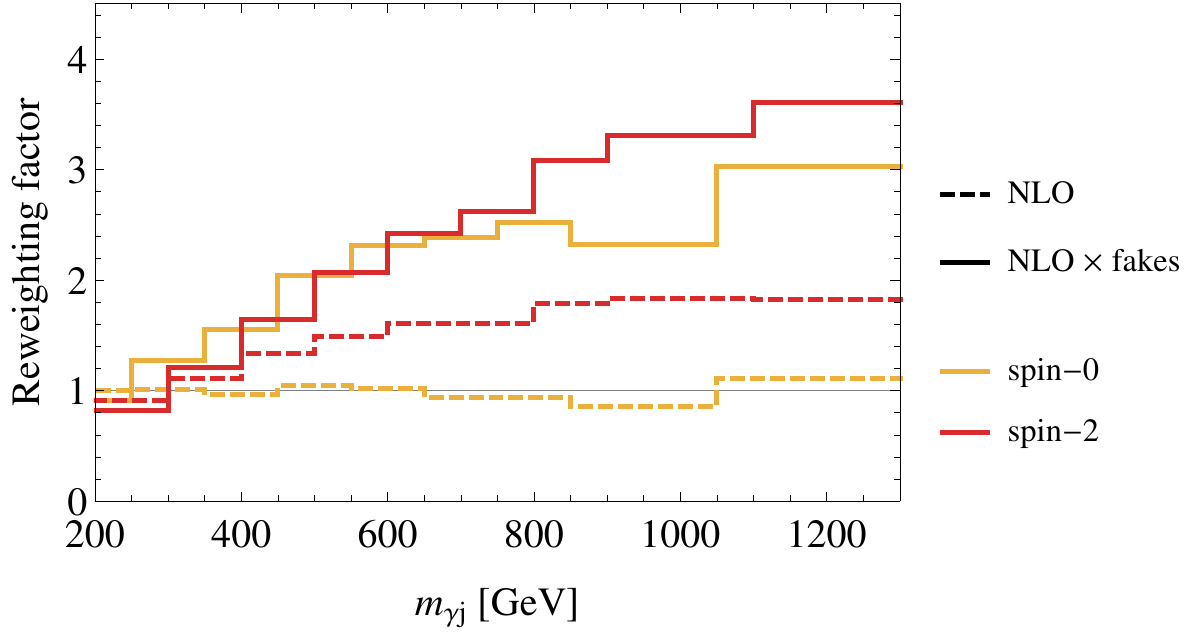}    
    \end{minipage}
  \end{subfigure}
  \begin{subfigure}[t]{0.50\linewidth}
    \begin{minipage}[t]{1.0\linewidth}
      \vspace{0pt}\includegraphics[width=\textwidth]{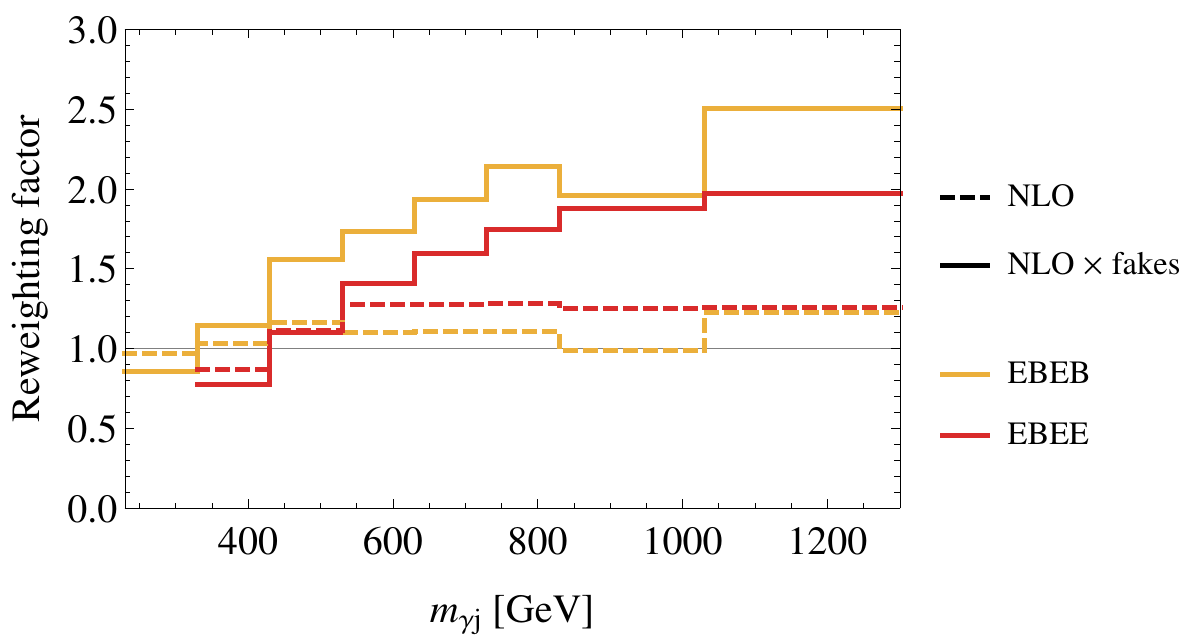}
    \end{minipage}
  \end{subfigure}
  \caption{Reweighting factor for $w_{\gamma j}$ with respect to the LO Monte Carlo distribution as
    a function of the invariant mass for ATLAS (left) and CMS (right). The dashed lines show the
    reweighting factor to modify the LO parton distribution $w_{\gamma j}^\tn{MG}$ to the NLO shape, including the effects of
    hadronization. The solid lines include in addition the reweighting due to a phase space
    dependent fake rate.}
  \label{fig:reweightingfactor}
\end{figure}

First, we calculate a photon-jet mass dependent K-factor using MadGraph5\_aMC@NLO, showered and
hadronized with Pythia~\cite{Sjostrand:2006za} and jets clustered with an anti-$k_T$, $R=0.4$
algorithm~\cite{Cacciari:2008gp} using FastJet~\cite{Cacciari:2011ma}.  We note that in the NLO
distribution we only consider the hardest jet of the event and we do not record its flavor.  This
step may be potentially improved by the use of an IR-safe jet flavor definition,
see~\cite{Banfi:2006hf,Buckley:2015gua,Larkoski:2014pca,Bauer:2013bza}.  Next, in order to model the
dependence of the fake rate on the pseudo rapidity and the transverse momentum we use the following
simplified ansatz for the jet rejection $r(p_T,\eta)$:
\begin{equation}
 r(p_T,\eta) = {\textrm{max}} \left\{ \frac{r_0}{1+ p_T/p_T^0 + \eta/\eta^0}, r_{\textrm{min}}\right\}\,, 
\end{equation}
where the functional form is motivated by the kinematical dependence of the jet-rejection rates as
estimated by ATLAS~\cite{ATLAS:2011kuc} and the parameter values $p_T^0 = 30$\,GeV, $\eta^0 = 4$ are
chosen to reproduce the rejection rate ratios between the lowest and highest lying $\eta$ and $p_T$
bins within uncertainties. Finally, $r_0/r_{\textrm{min}}$ is fairly uncertain as estimates of
rejection rates at very high $p_T$ and $\eta$ are not publicly available, but reproducing
experimental purity estimates in the forward region~\cite{Khachatryan:2016hje} leads to values in
the wide range $r_0/r_{\textrm{min}} \in [3,12]$\,.  The resulting reweighting factors compared to
the LO partonic $m_{\gamma j}$ distribution obtained from MadGraph5, $w_{\gamma j}^{\tn{MG}}$, at
both steps applied successively ($w_{\gamma j}^{\textrm{NLO}}/w_{\gamma j}^{\textrm{MG}}$ and
$w_{\gamma j}^{\textrm{NLO}\,\times\,\textrm{fakes}}/w_{\gamma j}^{\textrm{MG}}$) are shown in
Fig.~\ref{fig:reweightingfactor}.  We observe that with our choice of fake rate parameters, the
largest reweighting factors close to 3 are obtained above $m_{\gamma j} > 800$\,GeV for the ATLAS
spin-2 cuts. However, all experimental categories are affected by a reweighting factor which is a
combination of NLO, hadronization and faking effects, and which increases with the photon-jet
invariant mass until it saturates at some point.  This suggests a simple functional form for the
effective photon-jet spectrum deformation which we discuss next.
 
\subsection{Effective shape deformation}
In our effective ansatz for the deformation of the photon-jet spectrum, we focus on the invariant
mass region above $m_{\gamma\gamma} > 500$\,GeV below which the experiments have sufficient
statistics to control the mass distributions and calibrate their analyses and choose a simple,
linear form
\begin{align}
  \label{eq:1}
  w_{\gamma j}(\mathcal{R})&=\frac{w_{\gamma j}^{\textrm{MG}}}{N(\mathcal{R})} \left[ 1+\mathcal{R}  \times
      \begin{cases}
        0 &  m_{\gamma j} < 500\,\GeV\\
        \frac{\left.w_{\gamma j}^\tn{MG}\right|_{m_{\gamma j}=500\,\GeV}}{w_{\gamma j}^\tn{MG}}-1 & 500\,\GeV \leq
        m_{\gamma j} \leq 760\,\GeV\\
        \frac{\left.w_{\gamma j}^\tn{MG}\right|_{m_{\gamma j}=500\,\GeV}}{\left.w_{\gamma
              j}^\tn{MG}\right|_{m_{\gamma j}=760\,\GeV}} -1 & 760\,\GeV < m_{\gamma j}
      \end{cases}
\right]\,,
\end{align}
to roughly account for an overall kinematic dependence of the fake rate, hadronization and higher
order effects $N(\mathcal{R})$ is an $\mathcal{R}$-dependent normalisation factor. We define
$w_{\gamma j}(\mathcal R)$ such that $w_{\gamma j}(\mathcal{R}=0)$ corresponds to the partonic LO
distribution. Choosing $\mathcal{R}=1$, the shape of $w_{\gamma j}$ is unmodified for
$m_{\gamma j}<500\,\GeV$, then flat up to $m_{\gamma j}=760\,\GeV$, just above the observed peak of
the apparent excess, and finally is rescaled by the ratio of the differential cross sections at
$m_{\gamma j}=500\,\GeV$ and $m_{\gamma j}=760\,\GeV$ for $m_{\gamma j}>760\,\GeV$. The
$\mathcal{R}$-dependence of $N(\mathcal{R})$ is chosen such that the integral over $w_{\gamma j}$ is
always 1, independent of the value of $\mathcal{R}$. For the ATLAS spin-2 and the two CMS analyses,
the intervals in the above equation are shifted by $10\,\GeV$ to larger values due to the different
binning in these searches.\par

\begin{figure}[tb]
  \centering
  \begin{subfigure}[t]{0.55\textwidth}
    \begin{minipage}[t]{\textwidth}
      \vspace{0pt}\includegraphics[height=141pt]{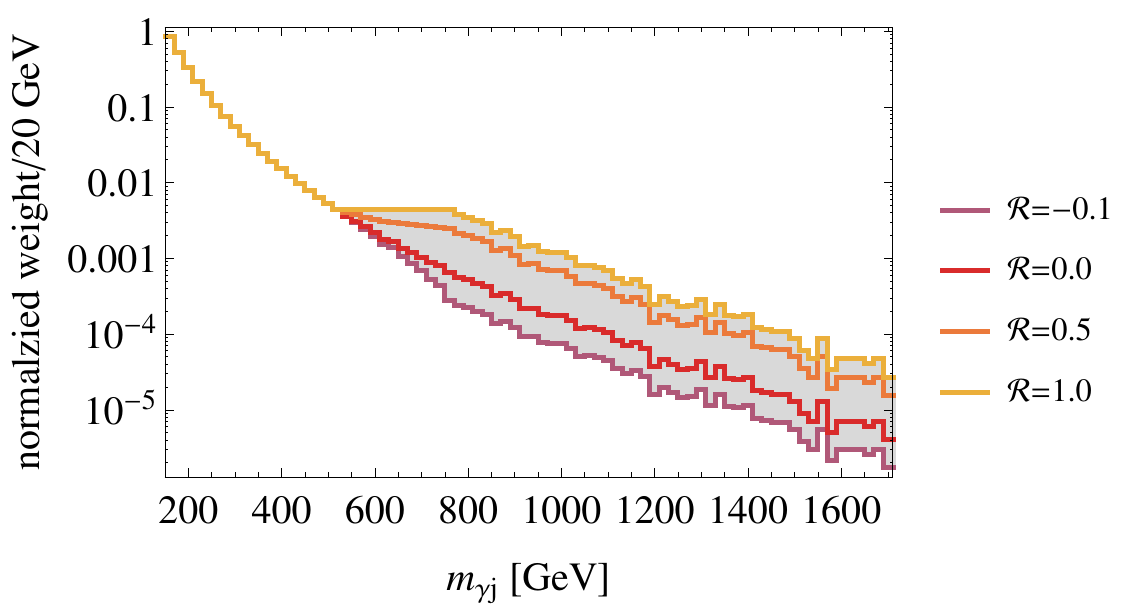}
    \end{minipage}
    \caption{}
    \label{fig:NormalizedDists}
  \end{subfigure}
  \begin{subfigure}[t]{0.42\textwidth}
    \begin{minipage}[t]{1\textwidth}
      \vspace{1pt}\includegraphics[height=136pt]{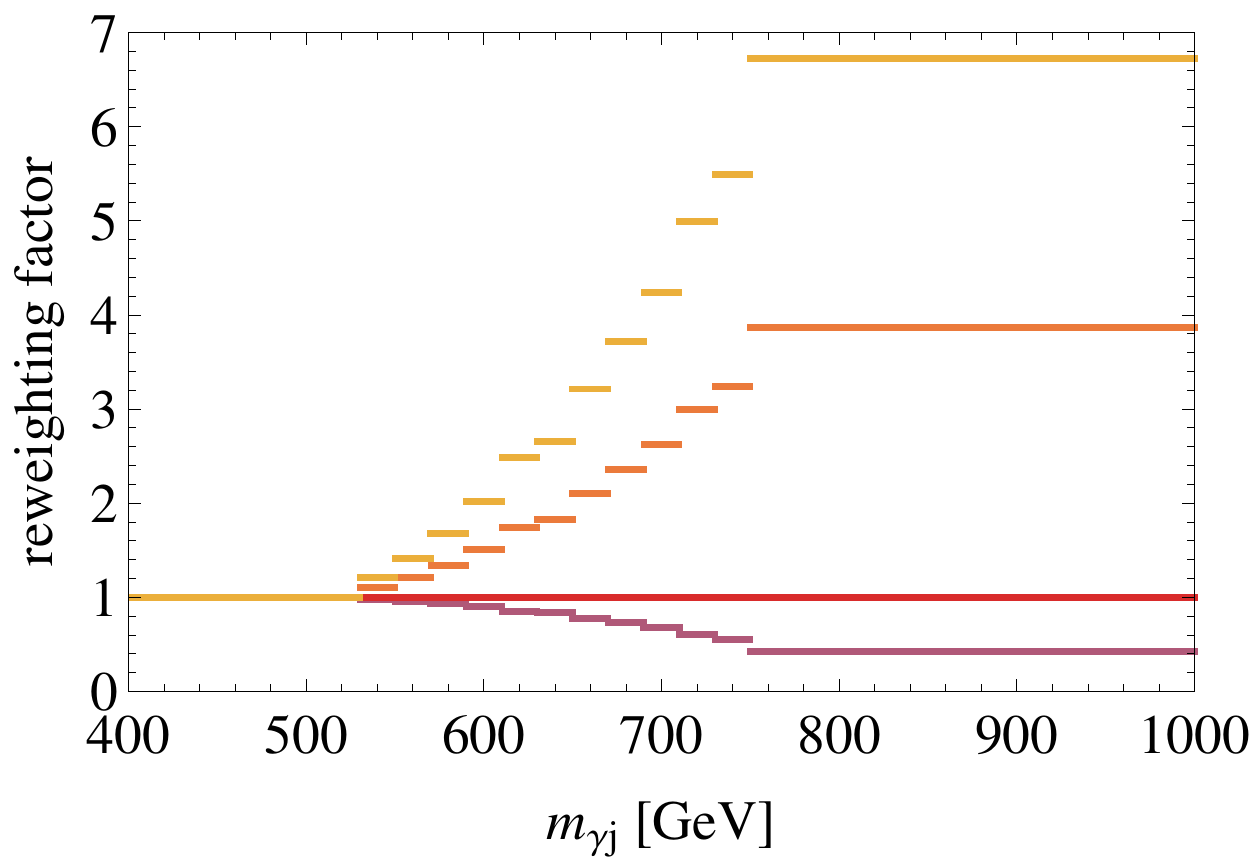}
    \end{minipage}
    \caption{}
    \label{fig:reweightingfunction}
  \end{subfigure}
  \caption{Left: Normalized invariant mass distributions after the ATLAS spin-0 cuts for the
    $p p \to \gamma j$ background with several choices for the interpolating parameter $\mathcal{R}$.  Right:
    Corresponding reweighting factor.}
  \label{fig:ReweightATLAS0}
\end{figure}

In the left panel of Fig.~\ref{fig:ReweightATLAS0}, we show the normalized differential
$p p \to \gamma j$ cross section for the ATLAS spin-0 analysis as a function of the invariant mass
for several choices of $\mathcal{R}$. In the right panel, the $\mathcal{R}$-dependent reweighting
factor of Eq.~\eqref{eq:1} is shown.  Since the spin-0 analysis applies the strongest cuts on the
transverse momenta of the photon candidates ($0.4\,m_{\gamma\gamma}$ and $0.3\,m_{\gamma\gamma}$,
respectively) its distribution is the steepest. Thus the reweighting factor of this analysis is the
largest being almost 7 above 770\,GeV. The maximal reweighting factors for the other analyses are
just above 6. We verified that increasing the flat region by 20\,GeV has only a small impact on the
reported results.
 
\begin{figure}[tb]
  \begin{subfigure}[t]{0.58\linewidth}
    \centering
    \vspace{0pt}\includegraphics[height=130pt]{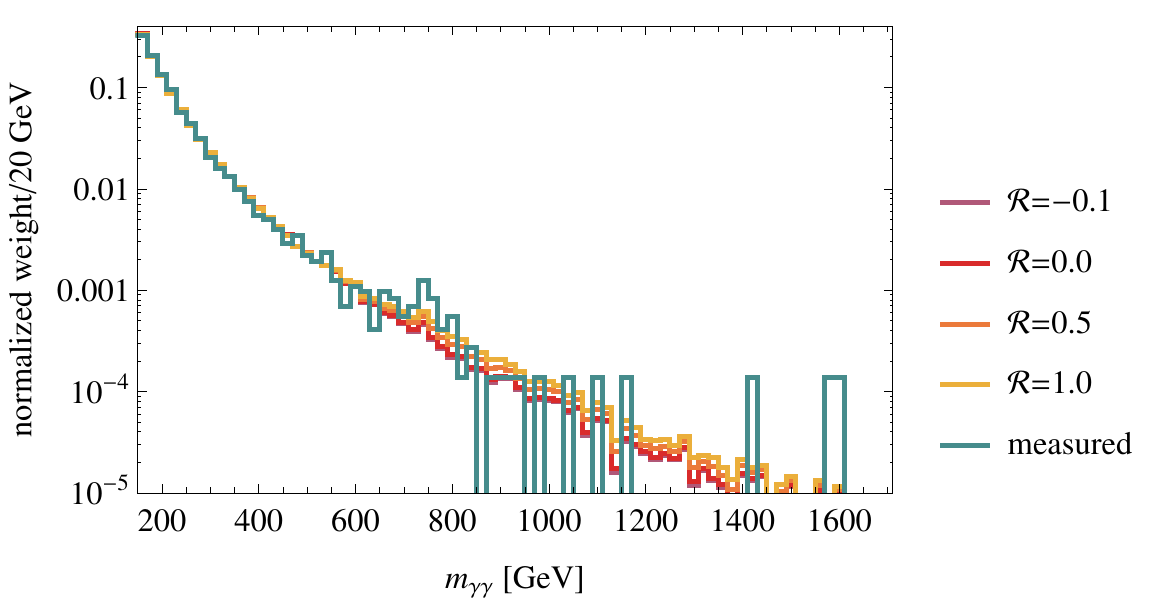}
    \caption{ATLAS spin-0}
    \label{fig:MixedDists}
  \end{subfigure}
  \begin{subfigure}[t]{0.4\linewidth}
    \centering
    \vspace{0pt}\includegraphics[height=130pt]{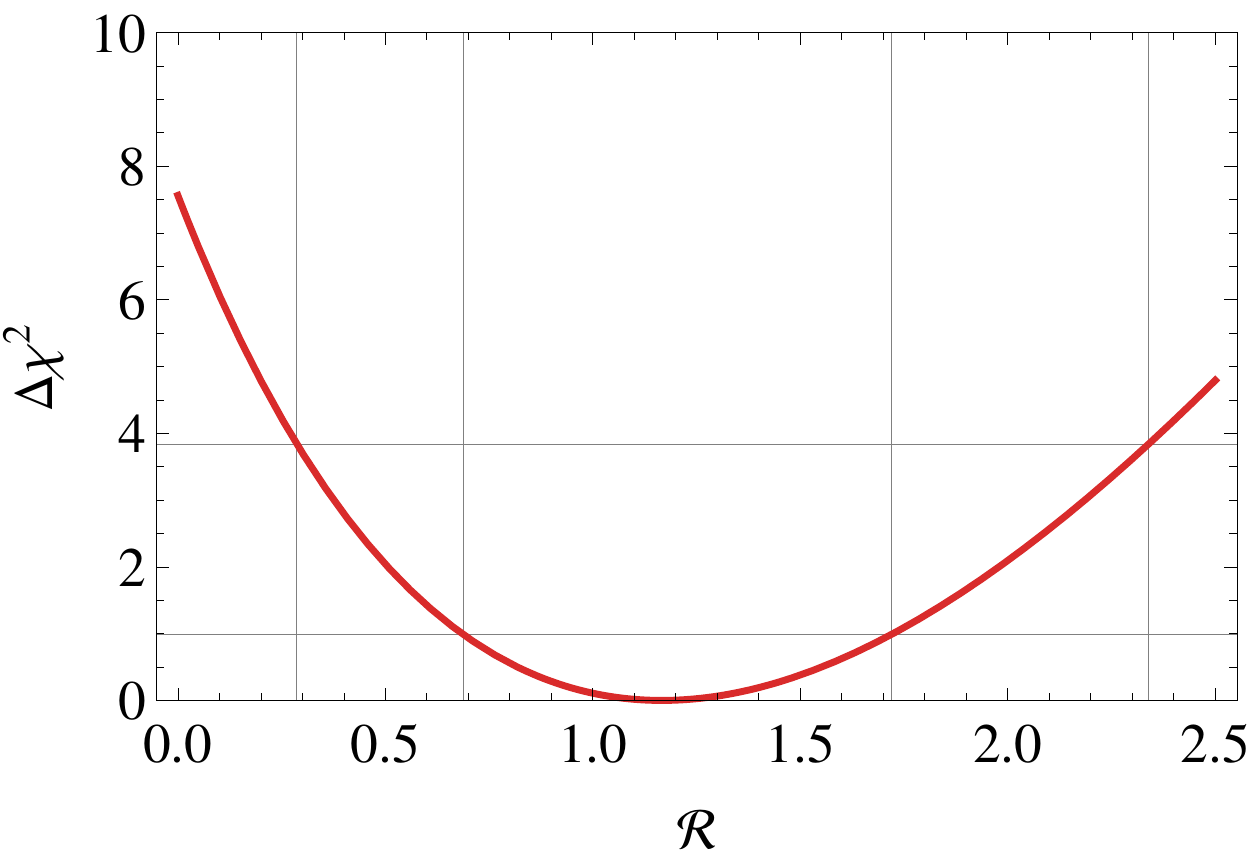}
    \caption{ATLAS spin-0}
    \label{fig:chisquared}
  \end{subfigure}
  \begin{subfigure}[t]{0.6\linewidth}
    \centering
    \vspace{0pt}\includegraphics[height=132pt]{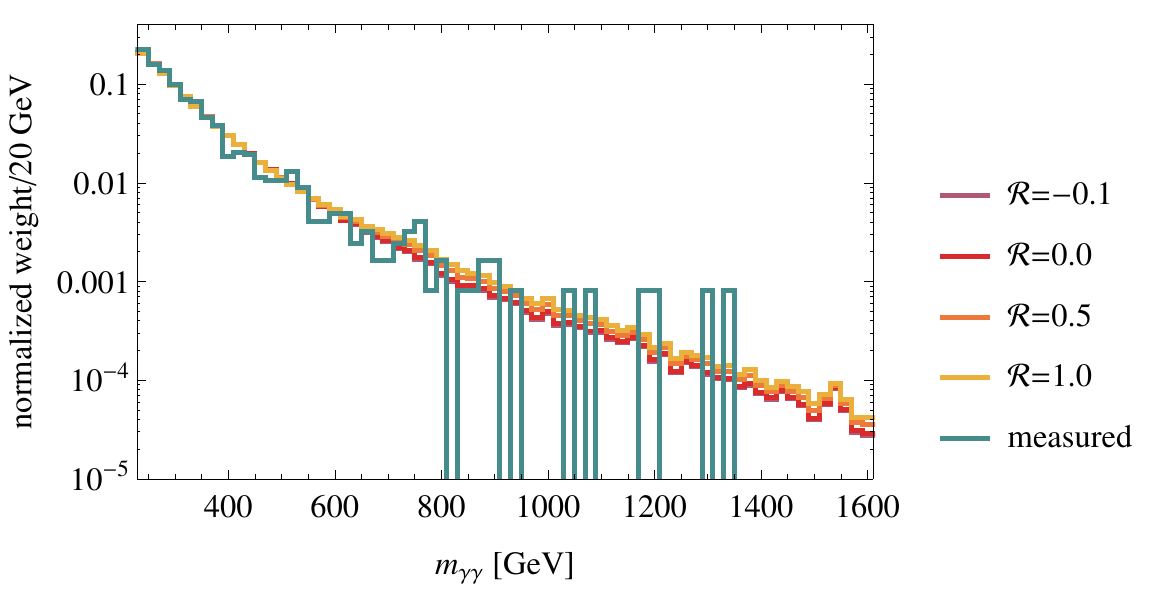}
    \caption{CMS EBEB}
    \label{fig:1MixedDists}
  \end{subfigure}
  \begin{subfigure}[t]{0.39\linewidth}
    \centering
    \vspace{0.8pt}\includegraphics[height=130pt]{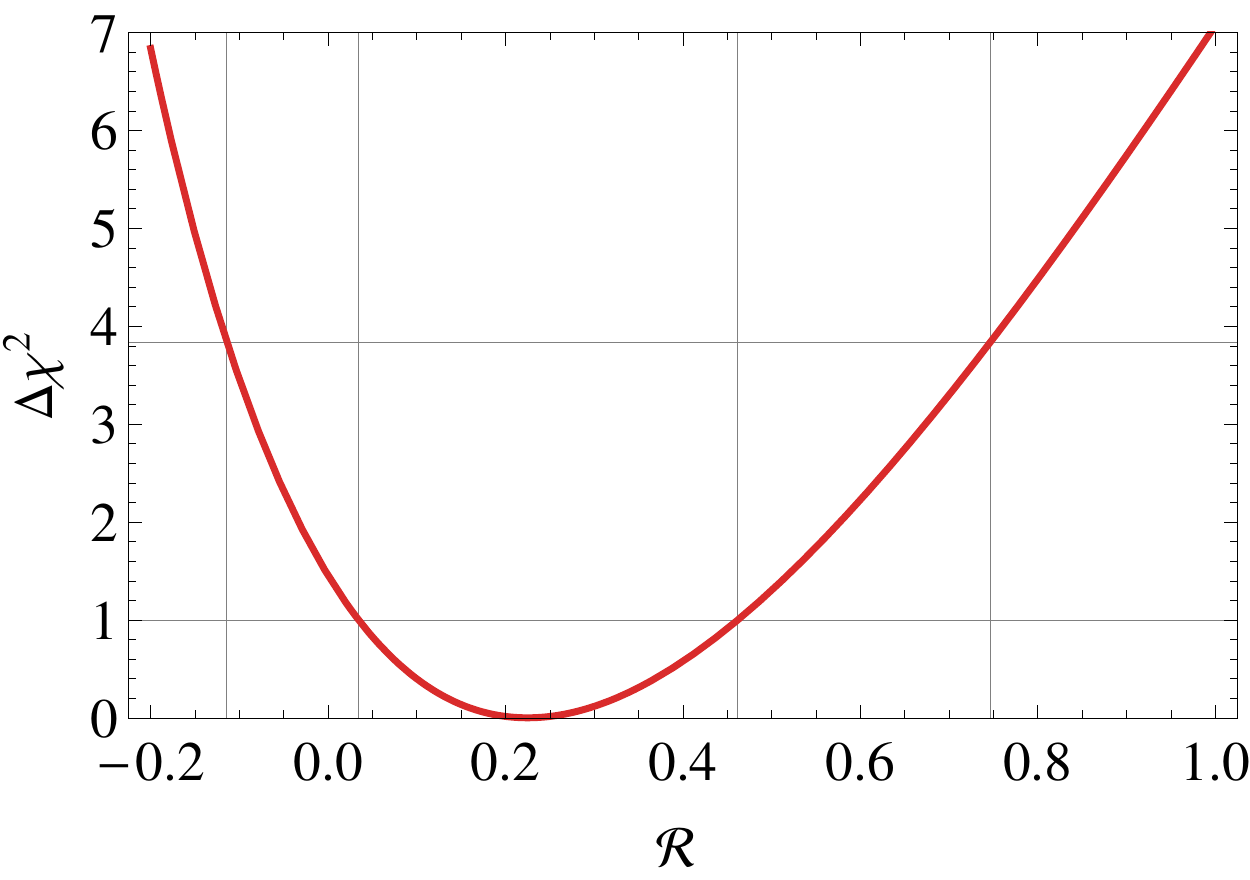}
    \caption{CMS EBEB and EBEE combined}
    \label{fig:CMSchisquared}
  \end{subfigure}
  \caption{Top left: Combined distributions $w_\tn{mix}$ for the ATLAS spin-0 cuts, all with an
    overall purity of 90\%. The distribution obtained from the 2015 data sets is shown in
    blue. Bottom left: Corresponding distribution for the CMS EBEB analysis. Top right:
    $\Delta\chi^2$ of fit to the ATLAS spin-0 distribution as a function of $\mathcal{R}$. The 1-
    and 2-$\sigma$ regions are indicated by the thin lines. Bottom right: corresponding plot for the
    combined CMS EBEB and EBEE analyses.}
\end{figure}

In Fig.~\ref{fig:MixedDists} and Fig.~\ref{fig:1MixedDists}, we show the resulting invariant mass
distributions $w_{\tn{mix}}$ for the ATLAS spin-0 and CMS EBEB analysis, respectively, on top of the
normalized distribution as measured in the 2015 data set.
\par
\begin{figure}[tb]
  \centering
  \begin{subfigure}[c]{0.49\linewidth}
    \centering
    \begin{minipage}[t]{1\textwidth}
      \includegraphics[width=0.95\textwidth]{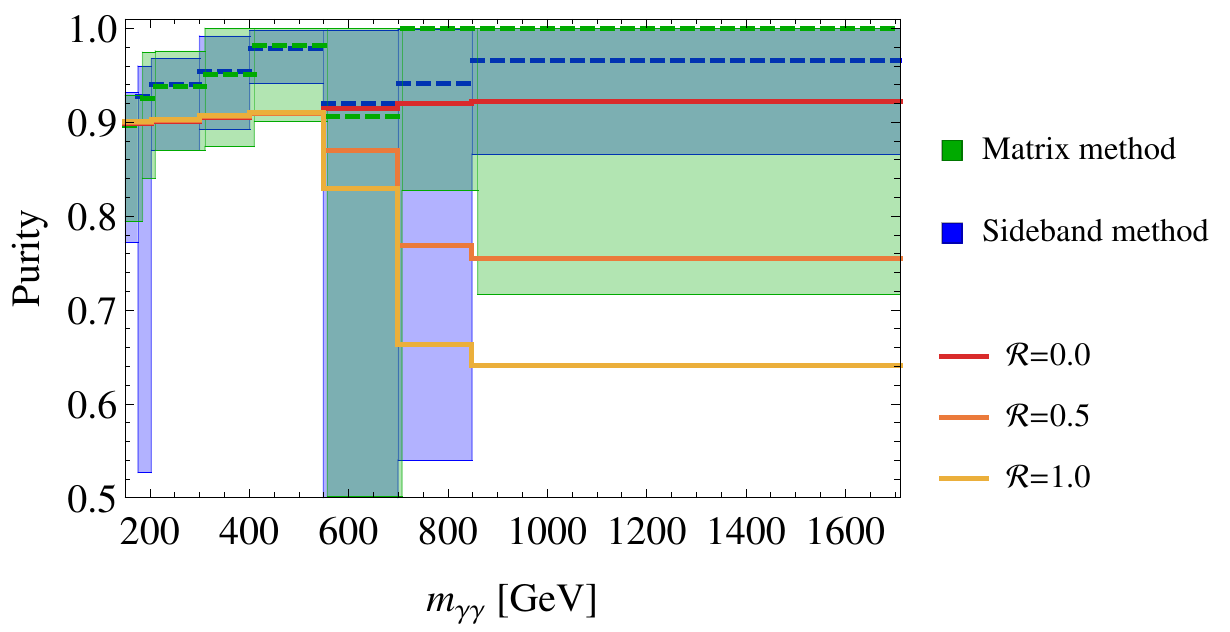}
    \end{minipage}
    \caption{ATLAS spin-0\qquad\qquad\quad\null}
    \label{subfig:Purity_0}
  \end{subfigure}
  \centering
  \begin{subfigure}[c]{0.49\linewidth}
    \centering
    \begin{minipage}[t]{1\textwidth}
      \vspace{1pt}\includegraphics[width=\textwidth]{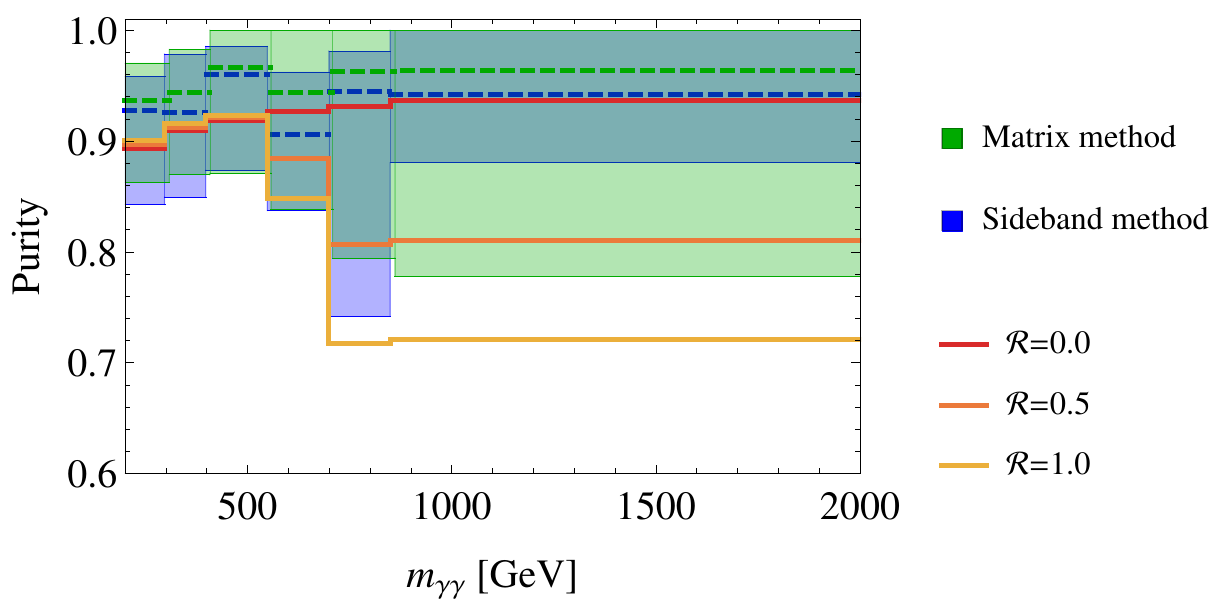}
    \end{minipage}
    \caption{ATLAS spin-2\qquad\qquad\quad\null}
    \label{subfig:Purity_2}
  \end{subfigure}

  \vspace{20pt}
  \centering
  \begin{subfigure}[b]{0.49\linewidth}
    \centering
    \includegraphics[width=\textwidth]{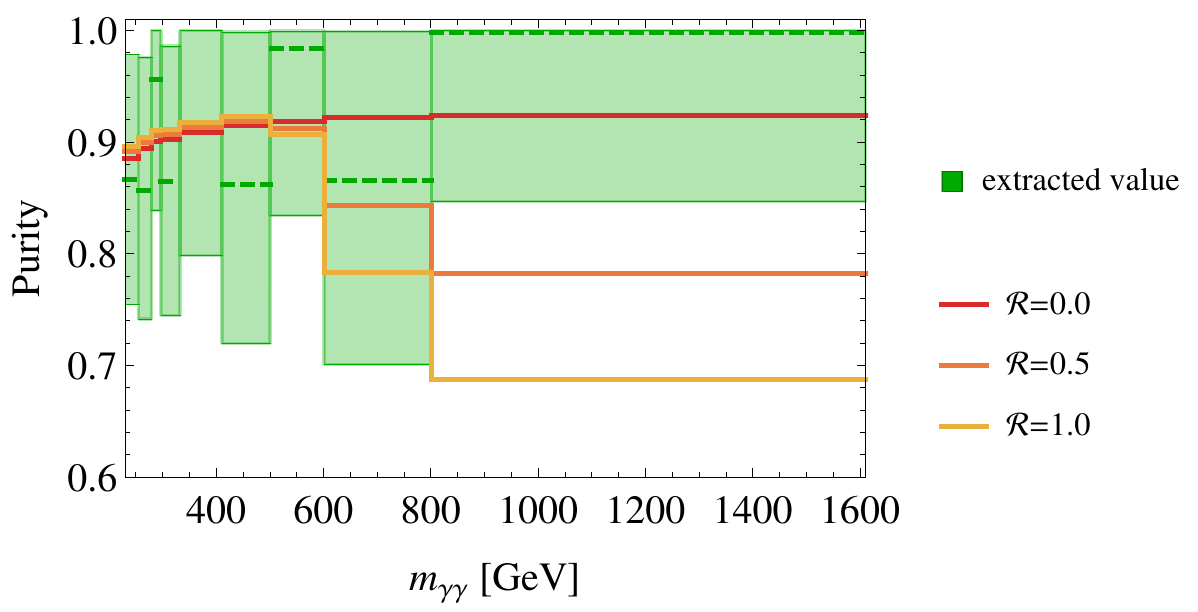}
    \caption{CMS EBEB\qquad\qquad\quad\null}
    \label{subfig:Purity_1}
  \end{subfigure}
  \centering
  \begin{subfigure}[b]{0.49\linewidth}
    \centering
    \includegraphics[width=\textwidth]{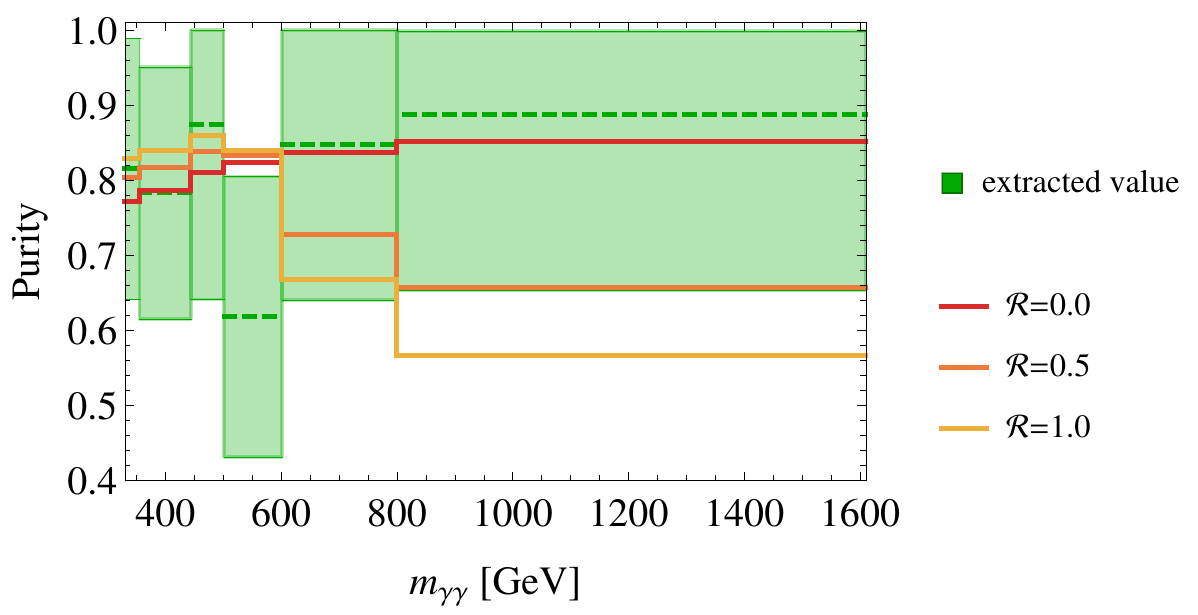}
    \caption{CMS EBEE\qquad\qquad\quad\null}
    \label{subfig:Purity_3}
  \end{subfigure}
  \caption{Purity of the combined distribution as a function of $m_{\gamma\gamma}$ for several
    choices of $\mathcal{R}$. The dashed lines show the central value for the purity as determined
    by the experiments and shaded the areas show the corresponding error.}
  \label{fig:purity}
\end{figure}

In addition to the average purity of the full sample, ATLAS and CMS try to estimate the purity as a
function of the diphoton invariant mass. This local purity is given by
\begin{equation}
  \label{eq:3}
  \mathcal{P}^i=\frac{\mathcal{P} w^i_{\gamma\gamma}}{\mathcal{P} w^i_{\gamma\gamma} + (1-\mathcal{P}) w^i_{\gamma j}}
\end{equation}
for the $i$-th bin. It can deviate significantly from the average purity $\mathcal{P}$ of the full
sample. In Fig.~\ref{fig:purity}, we show the binned purities for the mixed samples with several
choices of $\mathcal{R}$ compared to the purity determined by ATLAS with the $2\times 2$ sideband
\cite{Aad:2012tba} and the matrix method \cite{Aad:2011mh} and by CMS with a method described in
\cite{Chatrchyan:2014fsa}, respectively. We choose the same binning of the purity as is used in the
respective analysis.
\par
While the local purity is within the error band in most of the considered mass range (even for
$\mathcal{R}=1$), it does decrease for large invariant masses and our ansatz predicts a deviation
from the experimental value. Given the low statistics in this range, we consider this as a way to
falsify our proposal in the future rather than a contradiction with the currently available data.

\section{Statistical treatment}
\label{sec:stat-treatm}

The experimental analyses estimate the background shape by fitting a function $f(x)$ with
$x=m_{\gamma\gamma}/\sqrt{s}$ to the measured data. In the ATLAS spin-0 analysis, the following
ansatz is used:
\begin{align}
f(x)=N\left(1-x^{1/3}\right)^b x^{\sum_{j=0}^k a_j \left(\log x\right)^j}\label{ATLASfun}
\end{align} 
where $k=0$ was chosen. For the CMS analyses as well as the ATLAS spin-2, we use 
\begin{align}
f(x)=Nx^{a+b\log x}\,.\label{CMSfun}
\end{align} 
Note, that the ATLAS spin-2 analysis uses a mixture of Monte Carlo (for the $p p \to \gamma\gamma$
background) and data driven distributions (for the $p p \to \gamma j$ and $p p \to jj$ background),
leading to similar results as the fit function approach. In the data driven method, the shape of the
different backgrounds is obtained by extracting the corresponding events from control samples and
fitting their distribution with a function. The relative contribution to the observed
$pp \to \gamma\gamma$ sample is extracted from the data between
$200\,\GeV< m_{\gamma\gamma}<500\,\GeV$. For more details on this method see
\cite{Aaboud:2016tru}. Given the small statistics in the large invariant mass bins this approach
roughly corresponds to our LO MG distribution.\par

In order to see how the significance of the 750\,GeV excess changes with our ansatz, we fit the
distribution $w_{\textrm{mix}}$, defined in Eq.~\eqref{eq:5}, once with $w_{\gamma j}^{\textrm{MG}}$
corresponding to $\mathcal R=0$, and then with $w_{\gamma j}^{}$ as estimated at NLO with showering
and hadronization, including fakes and finally with $\mathcal{R}$ as a free fit parameter (as well
as the appropriate fit function $f(x)$) to the measured data. As an additional template, one could
extend the fit function $f(x)$ by a modification similar to the one described in Eq.~\eqref{eq:1},
which we will however not do for the sake of simplicity. The fits are performed with two methods
which yield similar results.

Firstly, we maximize the likelihood
\begin{align}
L=\prod_{i=1}^{N_\tn{bins}} P_{N_\tn{e}^i}(N_\tn{m}^i) 
\end{align}
where the product goes over all bins and $P_{N_\tn{e}}(N_\tn{m})$ is the Poisson probability to
measure $N_\tn{m}$ events when $N_\tn{e}$ events are expected.

Secondly, we minimize
\begin{align}
\chi^2=\sum_{i=1}^{N_\tn{bins}} (N_\tn{m}^i-N_\tn{e}^i)^2/N_\tn{e}^i
\end{align}
where we rebin the data such that each bin contains at least 10 events in order for the $\chi^2$
distribution to provide a reasonable description of the statistical uncertainties, see
e.g.~\cite{Cowan:2010js}.  In both fit methods, the overflow of the experimental histograms is
treated as one single bin. The best-fit parameters determine the number of expected events
$N_\tn{e}$ in the signal region (SR).\par

Since we are mostly interested in the local significance of the 750\,GeV excess, the SR is chosen by
eye from the measured distribution with the aim to capture the excess. We obtain a $p$-value by
comparing the number of measured events in the SR $N_\tn{m}$ with $N_\tn{e}$ (more precisely:
calculating the Poisson probability to measure at least $N_\tn{m}$ events):
\begin{equation}
   \label{eq:6}
  p=\sum\limits_{n=N_\tn{m}}^\infty \frac{N_\tn{e}^n}{n!} e^{-N_\tn{e}}\,.
\end{equation}
Clearly this simple approach which does not use any signal modeling and relies on a discrete width
and position of the SR is far from perfect. Consequently, it is not surprising that the obtained
significances of the excess are smaller than the ones reported by the experiments, even when we use
the same fit functions. Instead of focusing on absolute values one should therefore rather consider
the \textit{reduction} of the significance that results from modifying the background. The results
of the fits are shown in Table \ref{tab:stat_results}.\par

\begin{table}[tb!]
  \centering
  \resizebox{\textwidth}{!}{
    \begin{tabular}[c]{|r|r|c|c|c|c|c|c|c|}
      \hline
      \multirow{3}{*}{}
      &Analysis           & \mc{ATLAS spin-0}         & ATLAS spin-2     & \mc{CMS EBEB}             &
                                                                                                       \mc{CMS EBEE}          \\\hhline{|=========|}
      \parbox[t]{2mm}{\multirow{3}{*}{\rotatebox[origin=c]{90}{meas.}}}
      &SR                 & 730-770\,GeV & 720-760\,GeV  & 720-780\,GeV    & \mc{710-770\,GeV}         & \mc{710-770\,GeV}      \\\hhline{~~-------}
      &$\int \mathcal{L}\,\tn{d}t$  &  $3.2\,\ifb$ & $15.4\,\ifb$&  $3.2\,\ifb$    &  $2.7\,\ifb$& $12.9\,\ifb$&  $2.7\,\ifb$& $12.9\,\ifb$ \\
      &$N_m$              & 15           &     33      & 40             &  12         &  24         & 21          &  53       \\\hhline{|=========|}
      \parbox[t]{2mm}{\multirow{4}{*}{\rotatebox[origin=c]{90}{fitfunction}}}
      &$-2\log L$         &  270         &     -       & 330            &  200        &  -          & 200         &   -       \\
      &$\sigma$           &  3.4         &     -       & 2.9            &  1.9        &  -          & 1.7         &   -       \\\hhline{~--------}
      &$\chi^2/n$         &  1.6         &     0.75    & 1.2            &  0.80       &  1.0        & 1.2         &     0.98   \\
      &$\sigma$           & $3.1_{-0.2}^{+ 0.2}$ & $1.2^{+0.2}_{-0.2}$  &$2.9_{-0.3}^{+ 0.3}$ 
                                                                                                     &$1.8_{-0.2}^{+ 0.3}$ & $-1.5^{+0.2}_{-0.3}$ &$1.6_{-0.3}^{+ 0.3}$ & $-1.3^{+0.2}_{-0.2}$ \\\hhline{|=========|}
      \parbox[t]{2mm}{\multirow{13}{*}{\rotatebox[origin=c]{90}{distribution}}}
      &$\mathcal{R}$: \hfill$-2\log L$  & 270 & 360    & 330            & 210         &  310        & 210         &     300    \\
      &$\mathcal{R}$      & 0.86        &      0.11    & 0.97           & -0.048      & -0.7        & 0.24        &     -0.12  \\
      &$\sigma$           & 2.2         &      0.0     & 2.0            & 1.5         & -1.2        & 1.4         &     -0.23   \\\hhline{~--------}
      &MG: \hfill$\chi^2/n$ & 1.5       &      0.76    & 1.3            & 0.78        &  1.6        & 1.1         &     1.1    \\
      &$\sigma$           &$3.0_{-0.0}^{+ 0.0}$ & $0.2^{+0.0}_{-0.0}$ & $3.4_{-0.1}^{+ 0.1}$
                                                                                                     &$1.4_{-0.1}^{+0.1}$ & $-2.6^{+0.1}_{-0.1}$  &$1.9_{-0.2}^{+ 0.2}$ & $-0.86^{+0.14}_{-0.14}$  \\\hhline{~--------}
      &NLO: \hfill $\chi^2/n$& 1.5      & 0.76         & 1.2            & 0.80        &  1.6        & 1.2         &     1.0    \\
      &$\sigma$           &$3.0_{-0.0}^{+0.0}$ & $0.3^{+0.0}_{-0.0}$ &$3.2_{-0.1}^{+0.1}$  &
                                                                                             $1.4_{-0.1}^{+0.1}$ &$-2.6^{+0.1}_{-0.1}$& $1.7_{-0.2}^{+0.2}$ & $-1.2^{+0.1}_{-0.1}$   \\\hhline{~--------}
      &NLO$\times$fakes: \hfill$\chi^2/n$& 1.5 & 1.4   & 1.1            & 0.92        & 2.0         & 1.2         &     1.2    \\
      &$\sigma$           &$2.7_{-0.0}^{+0.0}$ &$-0.3^{+0.0}_{-0.0}$  & $2.9_{-0.1}^{+0.1}$ & $1.2_{-0.1}^{+0.1}$ &$-3.0^{+0.1}_{-0.1}$ &$1.4_{-0.2}^{+0.2}$&$-1.7^{+0.1}_{-0.1}$    \\\hhline{~--------}
      &$\mathcal{R}$: \hfill $\chi^2/n $&  1.2 & 0.75  & 1.0            & 0.81        & 1.3         & 1.1         &     1.1    \\
      &$\mathcal{R}$      &$1.2_{-0.5}^{+ 0.6}$ & $0.2^{+0.2}_{-0.2}$  &$1.1_{-0.4}^{+ 0.4}$
                                                                                                     &$-0.15_{-0.39}^{+ 0.51}$ &$-0.6^{+0.2}_{-0.2}$ &$0.30_{-0.22}^{+ 0.29}$ &$-0.091^{+0.084}_{-0.074}$  \\
      &$\sigma$           &$2.0_{-0.4}^{+ 0.4}$ &$-0.2^{+0.4}_{-0.4}$ &$1.9_{-0.5}^{+ 0.5}$  &$1.5_{-0.4}^{+ 0.4}$ & $-1.3^{+0.4}_{-0.4}$ &$1.2_{-0.4}^{+ 0.4}$ &$-0.40^{+0.42}_{-0.42}$       \\
      &$p_\tn{F-test}$    &   0.021     & 0.20          & 0.0027        & 0.73        & 0.014        &0.19         &     0.30  \\\hline
    \end{tabular}
  }
  \caption{Results of the fits to the data of all four analyses. In the first block from the top the
    signal region is defined and the number of measured events in this region $N_m$ is given. The
    results of a likelihood- and a $\chi^2$ fit of the fit function (value of the maximal likelihood
    and minimal $\chi^2$, respectively, and the local significance of the $750\,\GeV$ excess) are
    given in the second block. Finally, the third block contains the results of a likelihood and
    $\chi^2$ fit of the background distributions described in Section \ref{sec:setup} to the
    data. When $\mathcal{R}$ was fitted its best-fit value and the corresponding local significance
    of the excess are given, otherwise just the significance. In the last line the result of the
    F-test, testing whether $\mathcal{R}$ should be used as fit parameter, is given. For the
    minimized $\chi^2$ the parameter $n$ is the difference of number of bins and fit parameters. The
    errors indicate the 1-$\sigma$ interval of the systematic uncertainty of the fit. Note that the
    results of the spin-0 analysis with $3.2\,\ifb$ are based on the analysis with looser photon
    identification as described in~\cite{Aaboud:2016tru}.}
  \label{tab:stat_results}
\end{table}

Finally, an F-test is performed to determine if the generalization of our mixed distribution with
$w_{\gamma j}^{\textrm{MG}}$ to the one with $w_{\gamma j}(\mathcal{R})$ given in Eq.~\eqref{eq:1}
is needed to describe the data.  This test investigates the improvement of a fit when the fit
function is extended by an additional parameter. For this purpose, a test statistic
\begin{align}
F=\frac{(\chi^2_1-\chi^2_2)/(n_1-n_2)}{\chi^2_2/n_2}
\end{align}
is calculated, where $\chi^2_{1,2}$ are the minimized $\chi^2$'s of the two fit functions, $n_{1,2}$
are the numbers of bins (27 for the ATLAS spin-0) minus the number of input parameters (3 vs. 4 for
the fitting function and 1 vs. 2 for our distribution), and the subscripts refer to the two fit
functions with 2 signifying the extended function.  Eventually the $p$-value is determined as
\begin{align}
p_\tn{F-test}=\int_F^{\infty} \digamma(x; n_1-n_2,n_2)\tn{d}x,
\end{align} 
with $\digamma$ being the Fisher distribution. An additional fit parameter is warranted if
$p_\tn{F-test}<5\%$, see~\cite{Aaboud:2016tru}.\par

We find that for the ATLAS searches, the F-test suggests that $\mathcal{R}$ should be included as a
fitting parameter. The probability of an accidental improvement due to $\mathcal{R}>0$ is only
$2.1\%$ (ATLAS spin-0) and $0.27\%$ (ATLAS spin-2). On the other hand, the CMS categories do not
prefer a significant non-zero $\mathcal R$, see Table \ref{tab:stat_results}.  Furthermore, as a
consistency check, we apply the F-test on the ATLAS fitting function for spin-0,
Eq.~\eqref{ATLASfun}, and find that adding a $k=1$ component to the function does not pass the
test. Hence, as mentioned in~\cite{Aaboud:2016tru} only the leading term of the function with $k=0$
is retained.  The above in conjunction with the results collected in Table 2 suggest that it is
possible that the basis of functions used in Eq.~\eqref{ATLASfun} is not sufficient to accommodate
the deformation of the distribution proposed by us (or at least not the first term in the functional
form).\par

In the $\chi^2$ fit of the constructed distribution with $w_{\gamma j}^{MG}$ we find similar results
for the local significance as with the $\chi^2$ fit of the functional approach. However, in
particular in the two ATLAS analyses, using $\mathcal{R}$ as an additional fit parameter reduces the
local significance of the 750\,GeV excess by 1-1.5 units. The fact that the reduction is stronger in
the spin-2 analysis corroborates our working assumption that the background description deteriorates
in the forward region. This is further supported by the observation that in the CMS EBEB analysis,
which collects only events with both photon candidates in the central region, no reduction in
significance is observed and the best fit value for $\mathcal{R}$ is even slightly negative. Only in
the CMS EBEE analysis where one photon candidate is in the forward region the significance is
reduced by fitting $\mathcal{R}$, albeit less than in the ATLAS analyses.\par

Since $\mathcal{R}>0$ flattens the distribution one might worry that the reduction in the local
significance is obtained by overshooting the measured distribution in the high invariant mass
region. By verifying that both the minimal $\chi^2$ and the maximal likelihood hardly change between
the functional and the distribution fit we show that this is not the case.\par

As a final exercise, we try to obtain a ``combined'' significance from the analyses of the 2015 data
set. Clearly a proper statistical combination cannot be done, since we neglect correlations between
the various analyses and also fit for a single universal value of $\mathcal{R}$. Realistically,
$\mathcal R$ is expected to be somewhat different for the different analyses since they cover
different regions of phase space. Nevertheless, since the naive combination in Eq.~\eqref{sigs}
suffers from similar issues we set them aside and proceed as follows. We sum the $\chi^2$'s of the
analyses included in the combination and fit for a common $\mathcal{R}$ while keeping the
normalizations as separate variables. By combining the two CMS analyses we obtain $\sigma=2.4$ (1.9)
for $w_{\gamma j}^{\textrm{MG}}$ (with $w_{\gamma j}(\mathcal{R})$, best fit $\mathcal{R}=0.22$) and
$\sigma=1.9$ with $w_{\gamma j}^{\tn{NLO}\times\tn{fakes}}$. A combination of the ATLAS analyses is
impossible since they are not independent. However, we can combine each of them with the two CMS
analyses and obtain for ATLAS spin-0 combined with CMS $\sigma=3.6$ (2.6 with
$w_{\gamma j}(\mathcal{R})$, best fit $\mathcal{R}=0.46$; 3.1 with
$w_{\gamma j}^{\tn{NLO}\times\tn{fakes}}$) and for ATLAS spin-2 combined with CMS $\sigma=4.2$ (2.8
with $w_{\gamma j}(\mathcal{R})$, best fit $\mathcal{R}=0.53$; 3.4 with
$w_{\gamma j}^{\tn{NLO}\times\tn{fakes}}$), where the significance numbers before the brackets are
obtained for $w_{\gamma j}^{\textrm{MG}}$.

\section{The new energy frontier: searches beyond 1 TeV}
\label{sec:post-ichep-update}

Around ICHEP 2016, ATLAS and CMS updated their analyses, now based on 15.4\,$\ifb$ and 12.9\,$\ifb$,
respectively. In the updated ATLAS spin-0 analysis \cite{ATLAS:2016eeo} and the CMS EBEB and EBEE
analyses \cite{CMS:2016crm} the large excess around $750\,\GeV$ vanished and no other significant
excesses were found. An update of the ATLAS spin-2 analysis has not been presented. While CMS
processed the data exactly as before, ATLAS made some adjustments, perhaps most importantly, using a
tighter photon isolation. We repeat the fits and the statistical treatment of the reported results
with the same methods as described above and report the results for the larger dataset in Table
\ref{tab:stat_results}. Note that there is a downwards fluctuation in the signal region in the full
CMS dataset which even leads to a slightly negative significance.\par
\begin{figure}[tb]
  \centering
  \begin{subfigure}[b]{0.45\linewidth}
    \centering
    \includegraphics[width=1\textwidth]{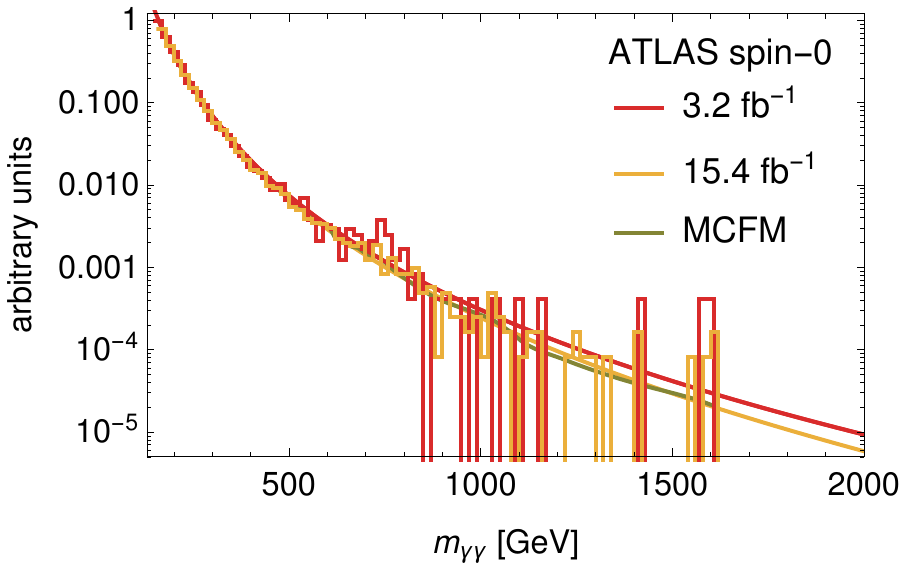}
    \label{subfig:comp_ATLAS}
  \end{subfigure}\hfill
  \begin{subfigure}[b]{0.45\linewidth}
    \centering
    \includegraphics[width=\textwidth]{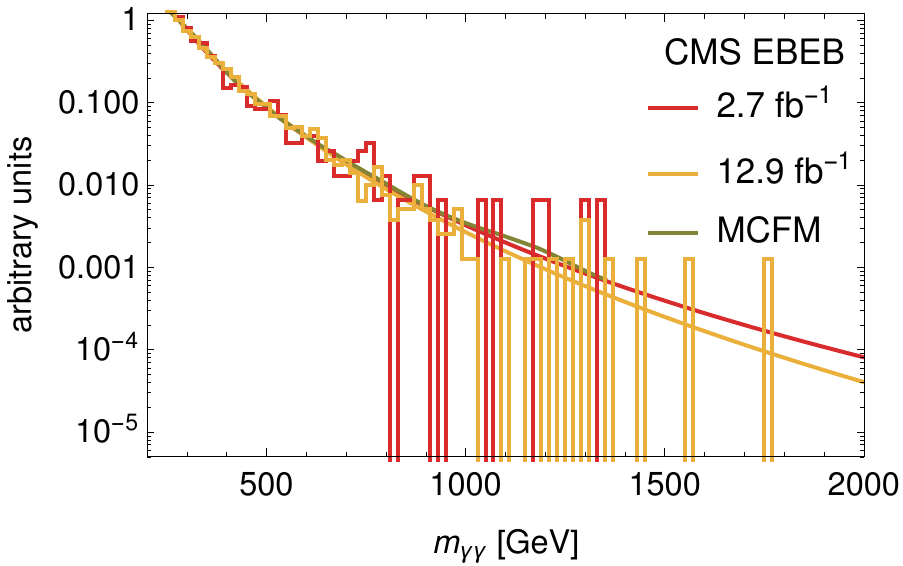}
    \label{subfig:comp_EBEB}
  \end{subfigure}\\
  \begin{subfigure}[b]{0.45\linewidth}
    \centering
    \includegraphics[width=\textwidth]{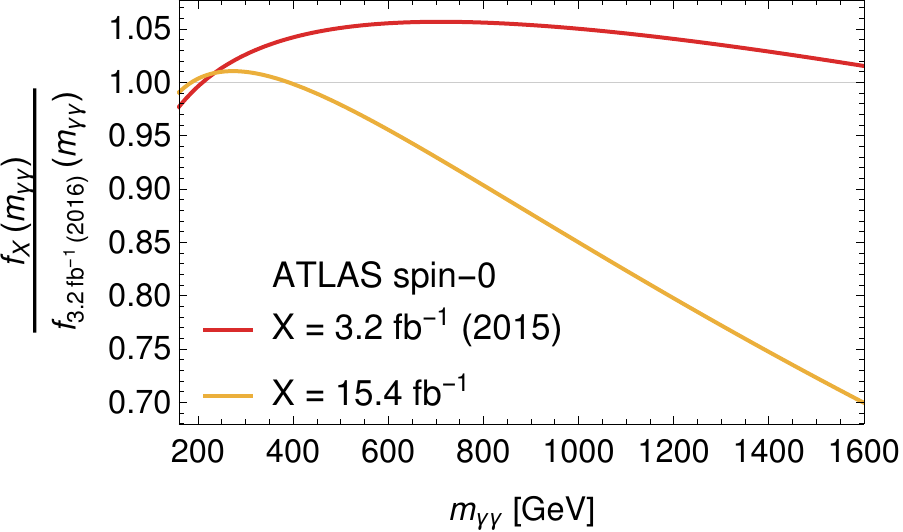}
    \label{subfig:small_oldvsnew}
  \end{subfigure}\hfill
  \begin{subfigure}[b]{0.45\linewidth}
    \centering
    \includegraphics[width=\textwidth]{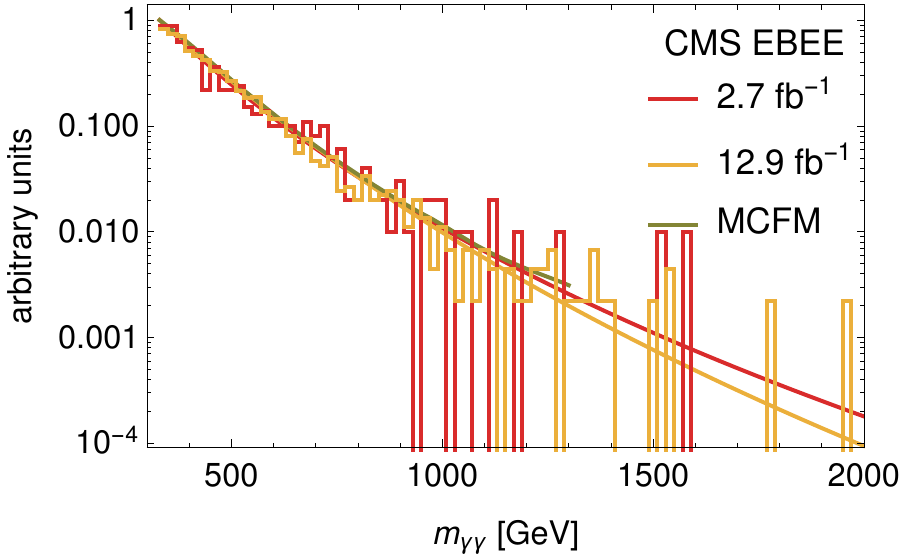}
    \label{subfig:com_EBEE}
  \end{subfigure}
  \caption{Upper plots and lower right plot: Comparison of the measured and fitted distributions
    with the small and the full datasets. In addition, the pure digamma spectrum as obtained from
    MCFM is shown. In the upper left plot the comparison is between the smaller Moriond 2016 dataset
    with the old photon isolation method and the full ICHEP 2016 dataset. The distributions and
    functions are normalized to have the same value at the low $m_{\gamma\gamma}$ end of the
    histograms.\\
    The lower left plot shows the ratios of the normalized fit functions $f_X$ fitted to the ATLAS
    spin-0 data set $X$ with the year 2015 (2016) in the brackets indicating the old Moriond
    (updated ICHEP) photon isolation criteria.}
  \label{fig:comparison}
\end{figure}

\begin{figure}[tb]
  \centering
  \begin{subfigure}[b]{0.32\linewidth}
    \centering
    \includegraphics[width=1\textwidth]{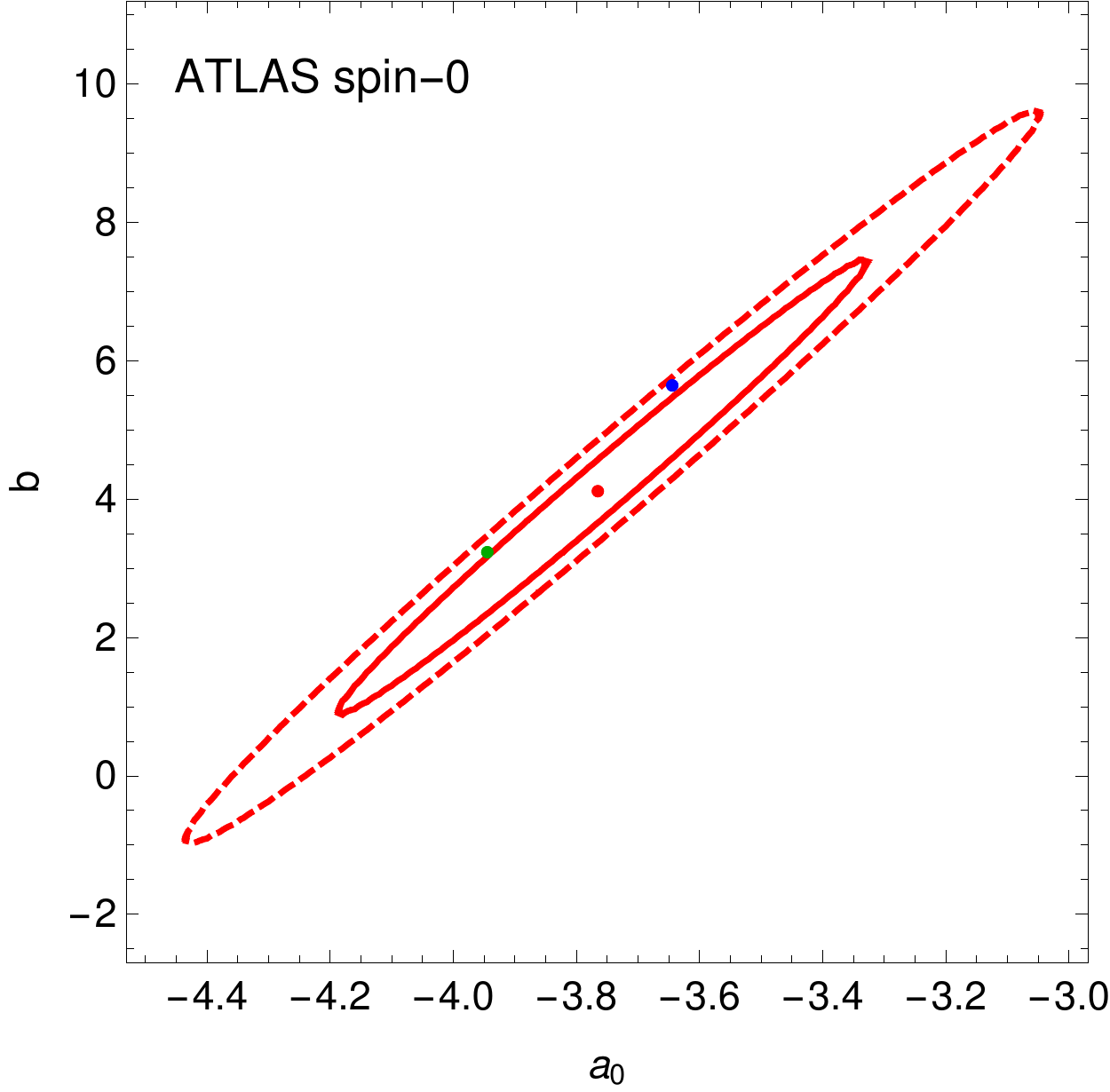}
    \label{subfig:fitcomp_ATLAS}
  \end{subfigure}\hfill
  \begin{subfigure}[b]{0.32\linewidth}
    \centering
    \includegraphics[width=\textwidth]{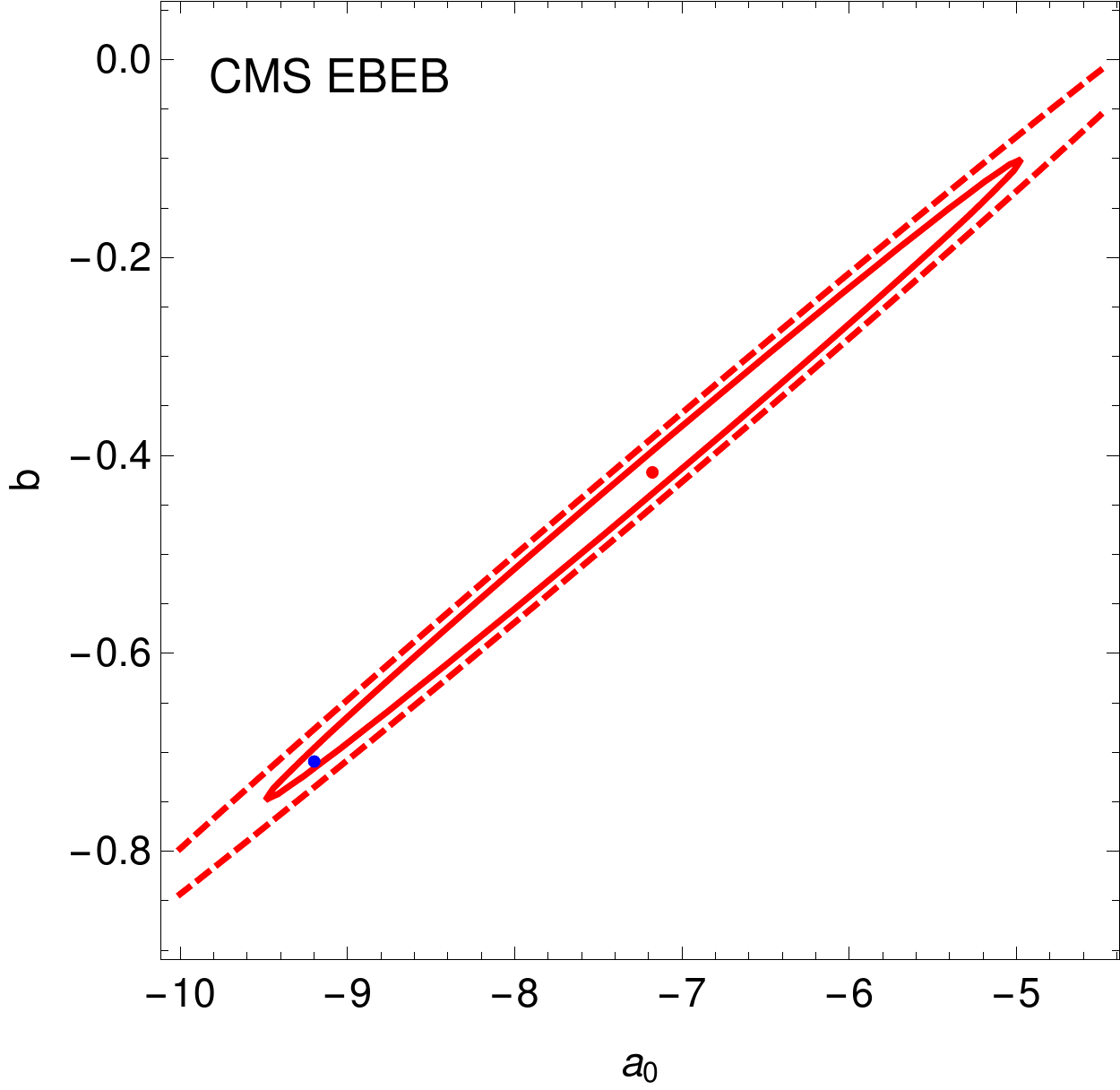}
    \label{subfig:fitcomp_EBEB}
  \end{subfigure}\hfill
  \begin{subfigure}[b]{0.32\linewidth}
    \centering
    \includegraphics[width=\textwidth]{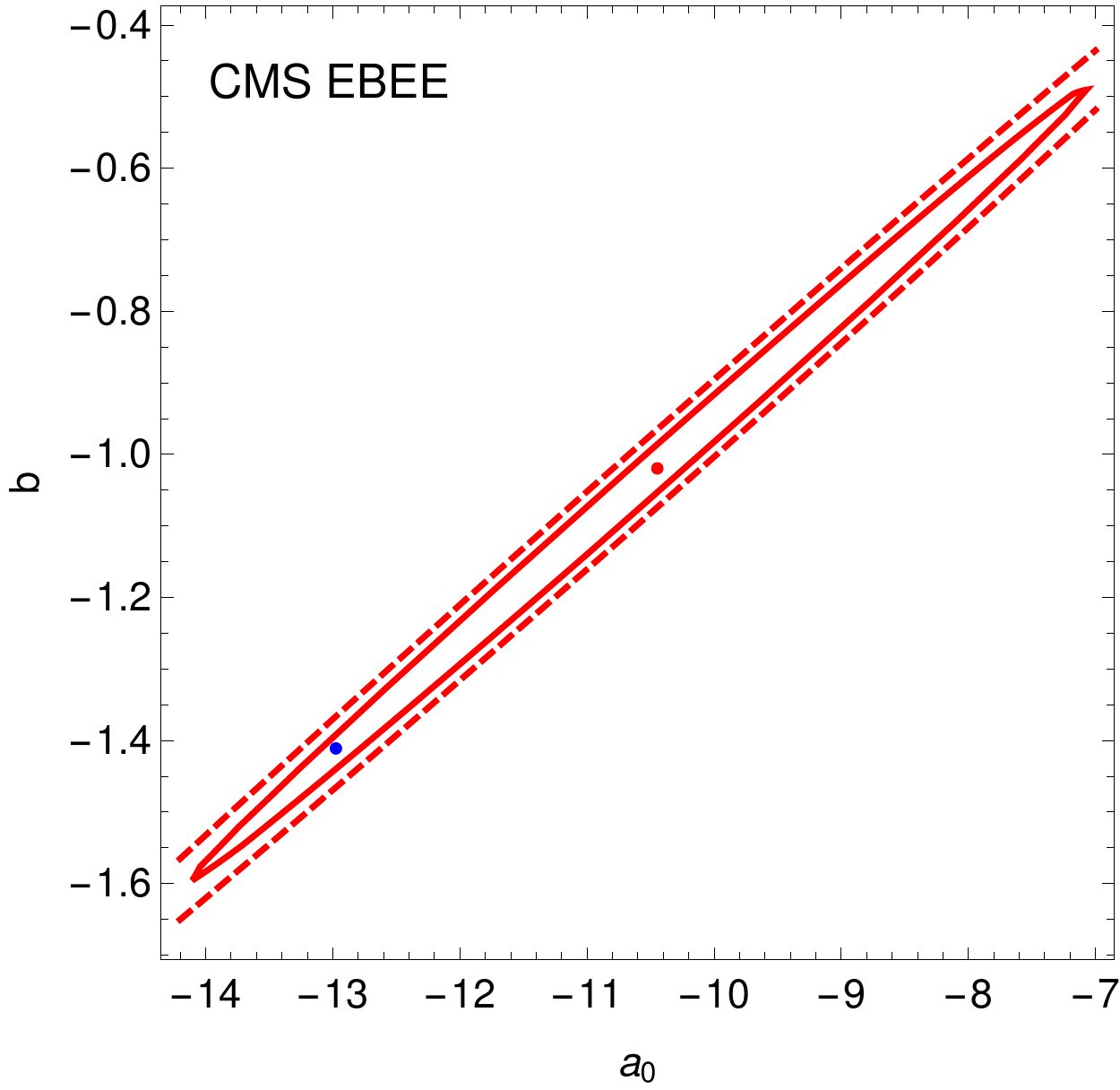}
    \label{subfig:fitcomp_EBEE}
  \end{subfigure}
  \caption{Best fit point of the $\chi^2$ fit of the appropriate function to the Moriond 2016
    dataset in red with the 1- and 2-$\sigma$ contours. The best fit point for the fit to the full
    ICHEP 2016 dataset is shown in blue and in the left plot the best fit point for the ATLAS spin-0
    $3.2\,\ifb$ dataset with the new photon isolation is shown in green.}
  \label{fig:fit_comparison}
\end{figure}

\begin{figure}[tb]
  \centering
  \begin{subfigure}[b]{0.5\linewidth}
    \centering
    \includegraphics[width=\textwidth]{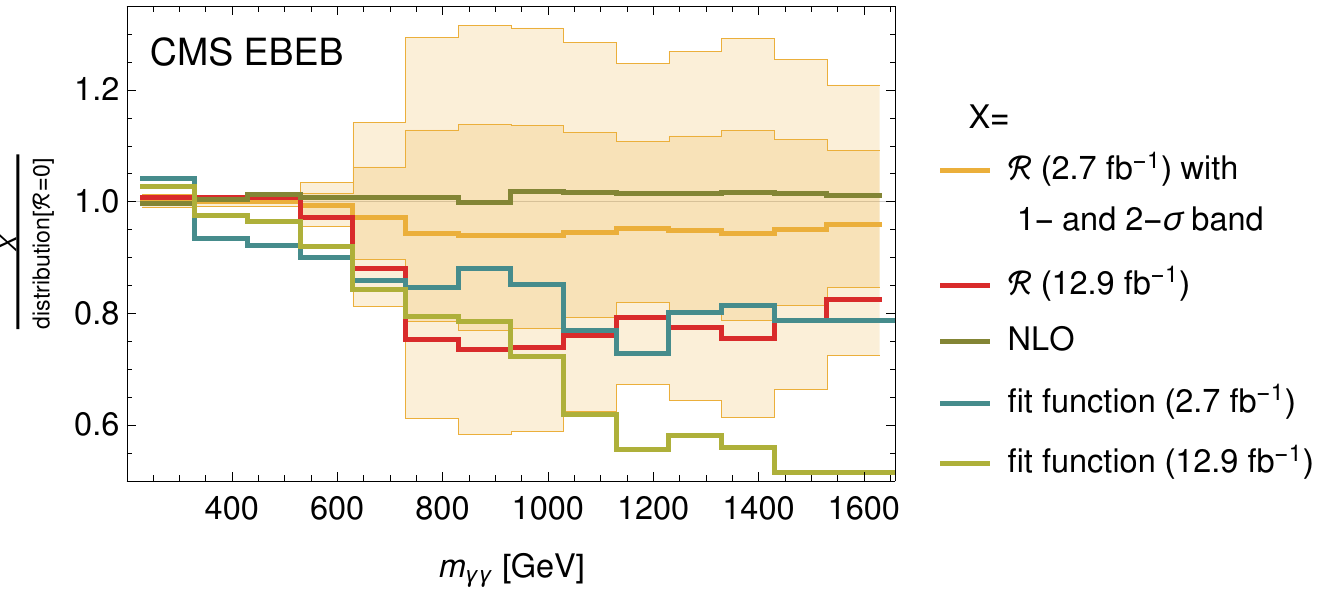}
    \label{subfig:ErrorBands_EBEB}
  \end{subfigure}\hfill
  \begin{subfigure}[b]{0.5\linewidth}
    \centering
    \includegraphics[width=\textwidth]{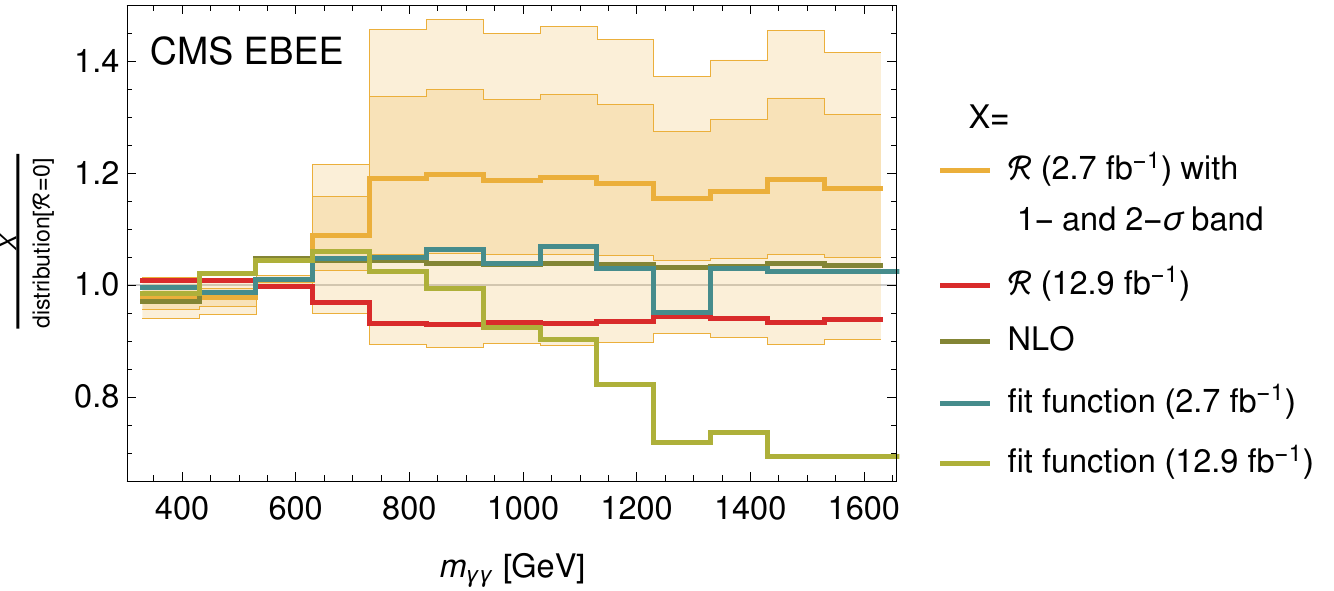}
    \label{subfig:ErrorBands_EBEE}
  \end{subfigure}\\
  \begin{subfigure}[b]{0.5\linewidth}
    \centering
    \includegraphics[width=1\textwidth]{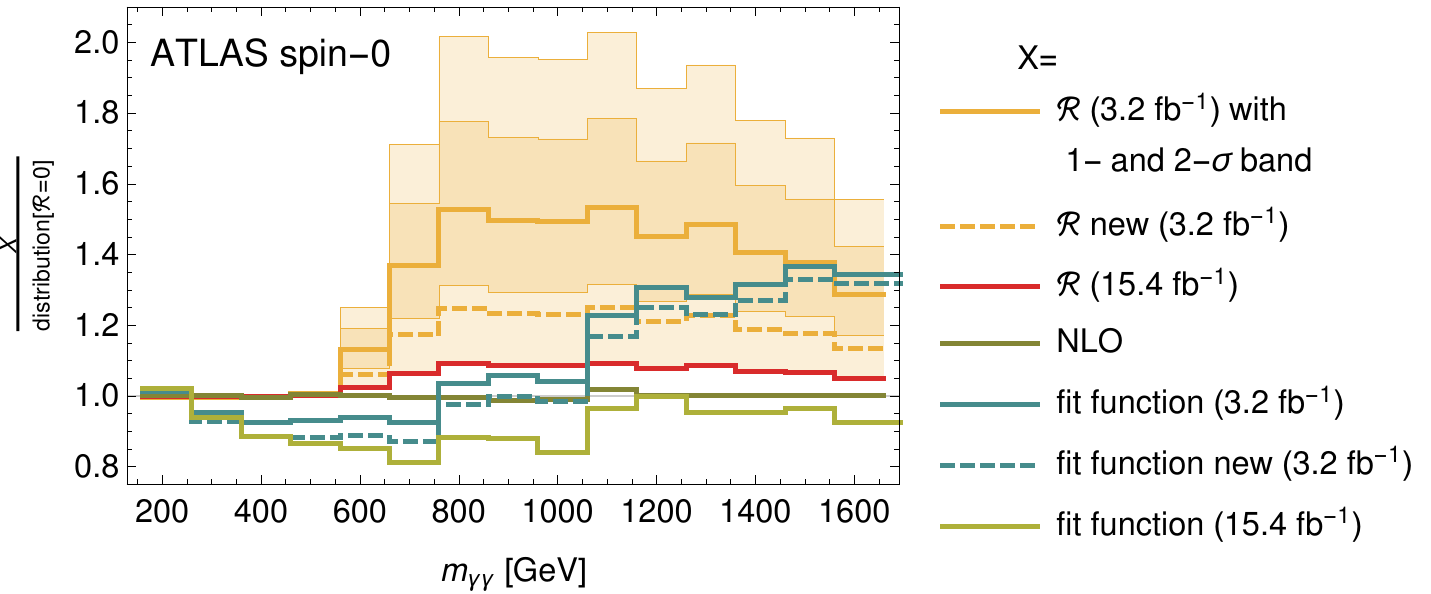}
    \label{subfig:ErrorBands_ATLAS}
  \end{subfigure}
  \caption{Plot of the normalized distributions with several choices of $\mathcal{R}$ and normalized
    fit functions to the old and new datasets, all divided by the distribution for
    $\mathcal{R}=0$. In the plot for the ATLAS spin-0 analysis the dashed lines show the results
    obtained using the old dataset with the new photon isolation criteria.}
  \label{fig:ErrorBands}
\end{figure}

Comparing the new fit functions to the ones based on the previous small datasets presented at
Moriond 2016 we find a steeper functional fit in all three analyses, see
Fig.~\ref{fig:comparison}. While the new best fit parameters are within one standard deviation for
the two CMS fits, the ones for the ATLAS fit deviate by almost two standard deviations after
marginalizing over the normalization, see Fig.~\ref{fig:fit_comparison}. This might, however, be an
effect of the changed photon isolation as the fit to the $3.2\,\ifb$ dataset with the updated photon
identification also deviates by more than one standard deviation from the previous best fit point. A
better understanding of the effect of the fake photons could be obtained by investigating the result
of changing the isolation criteria with the full $15.4\,\ifb$ dataset. The tighter isolation
criteria are also reflected in the better agreement between the fitted distributions and the MCFM
generated digamma spectrum.\par

In order to show the changes in the fits, the ratio of the normalized fit functions for the ATLAS
spin-0 analysis is shown in the lower left plot of Fig.~\ref{fig:comparison}. A direct comparison of
the data and the un-normalized fit functions, even for the fits to the two different $3.2\,\ifb$
sets, is difficult since the binning of data has changed. The large change in the fit parameters is
reflected in the deviation of more than 5\,\% for the comparison of the fits to the two $3.2\,\ifb$
datasets and the even greater deviation when comparing with the fit function to the full
$15.4\,\ifb$ dataset. While in the previous signal region near 750\,GeV the change is of the order
of 10\,\% it is greater than 30\,\% near 1.6\,TeV. This shows that the actual shape of the digamma
spectrum at high invariant masses is hard to predict precisely by an extrapolation and is therefore
very much subject to systematic uncertainties.\par

Finally in Fig.~\ref{fig:ErrorBands} the ratios of several normalized distributions and fit
functions to the normalized distribution with $\mathcal{R}=0$ are shown. These include the
distributions with the best fit value for $\mathcal{R}$ based on the Moriond 2016 dataset and the
ICHEP 2016 dataset and also the NLO distributions and the fit functions to the old and new
datasets. In the case of the ATLAS spin-0 analysis also the distribution and fit function to the
2015 dataset with the new photon isolation is shown. By comparing the curves we find that a sizable
systematic uncertainty can be inferred from the differences between the fit functions.

\section{Conclusions}
\label{sec:conclusions}

This paper deals with a problem that often arises in searches for new physics at the energy
frontier. In this context the challenge is to look for a new resonance at the upper end of a
distribution where only limited knowledge on the SM background is available. As a case study we
focus on the 750\,GeV anomaly where we examine in particular the implications of the possibility
that the excess in the 2015 data set is not only due to a (malicious) statistical fluctuation but
also a result of a physical effect. We discuss possible issues with the background: how much
photon-jet contamination is still allowed in the region of interest?  How could it affect the
significance of the excess?

We study these questions using currently available theoretical tools for computing the photon-jet
mass distributions and apply them to the small set of publicly available data. However, this
approach is limited by our ability to thoroughly disentangle the effects of the
$(p_T,\eta)$-dependent jet fake rate and the theoretical uncertainty of the shape of the photon-jets
background. We therefore choose to model these combined effects by an $m_{\gamma j}$ dependent
reweighting of the invariant mass distribution, keeping the overall purity within the quoted ranges.
We first study a physics-driven reweighting procedure: we convolve a mass dependent K-factor with a
rapidity and transverse momentum dependent photon fake rate for the jets. The K-factor is extracted
comparing the NLO leading-jet-photon to the LO quark-photon spectrum, and the phase space dependence
of the fake rate is estimated from the experimental literature~\cite{ATLAS:2011kuc}. Both correction
factors are approximate, based on incomplete information, and should be taken with a grain of
salt. Motivated by this result, we then consider a more phenomenological deformation of the
$p p \to \gamma j$ spectrum. It allows us to study the sensitivity of the significances on a single
continuous quantity $\mathcal{R}$ (see Eq.~\eqref{eq:1}) which parametrizes an effective
deformation.
 
To summarise our results for the 750\,GeV case study based on the 2015 data we focus on the simpler
effective ansatz where we find the following:
\begin{itemize} 
\item For the ATLAS spin-0 analysis, the significance of the excess can be reduced by
  $\Delta \sigma \simeq1.1 $ when comparing the fitting function defined in \eqref{ATLASfun} with
  our best-fit to the $\mathcal R$-modified distribution.  A comparable reduction is found for the
  ATLAS spin-2 measurement. Here however, it is less straightforward to determine the reduction,
  since the estimation of the background shape in our ansatz differs from that of the ATLAS
  analysis, which is not reproducible since the required data is not publicly available. Strictly
  speaking, $w_{\gamma j}^{\textrm{MG}}$ does therefore not correspond to the ATLAS approach but is
  the best approximation we can get. Since, however ATLAS claims to find comparable results with the
  corresponding fitting function defined in \eqref{CMSfun}, we can reduce the significance with the
  $\mathcal{R}$-modified distribution with respect to the fitting function by
  $\Delta \sigma \simeq 1.0$ as well as with respect to the distribution with
  $w_{\gamma j}^{\textrm{MG}}$ by $\Delta\sigma\simeq 1.5$.
\item The effect is smaller for the CMS 13\,TeV analyses with $\Delta \sigma \simeq 0.3-0.4$,
  depending on the category.
\item In a ``combined fit'' to independent ATLAS and CMS datasets, the significance can be reduced
  by as much as $\Delta \sigma \approx 1.0 \,(1.4)$ for the ATLAS spin-0 (spin-2) combined with CMS.
\end{itemize}

The larger preference for an enhanced photon-jet contribution in the spin-2 sample could point to
its higher sensitivity to the large rapidity region where jet fakes are more difficult to
reject. Finally, an F-test shows that the ATLAS data support using a more complex distribution.

To summarise our results for the 750\,GeV case study based on the 2016 data we find the following:
\begin{itemize}

\item For the ATLAS spin-0 analysis, we find that the new data prefers $\mathcal{R}$ in the range
  $0.2\pm 0.2$, eliminating the remaining significance of $1.2\,\sigma$ in the full data set. As for
  the spin-2 case no data is currently available.
  
\item The updated CMS analyses based on $12.9\,\ifb$ even have a downwards fluctuation with respect
  to the fit function near 750\,GeV leading to negative significances. Correspondingly the best fit
  values for $\mathcal{R}$ are negative and ameliorate the situation.
  
\end{itemize}

We emphasize that our simplified ansatz for the effective modification of the photon-jet background
with $\mathcal{R}$ is not meant to necessarily represent a new background source nor the exact shape
of the background contamination in the signal region. Rather its envelope (corresponding to the
shaded area in Fig.~\ref{fig:NormalizedDists}) is expected to reflect a possible combination of
higher order QCD contributions, fragmentation, isolation and detector effects, which are outside of
theoretical control and, in the high invariant mass signal region, also beyond direct experimental
probes with currently available data. We further note that a quark flavor tagged dijet sample might
provide a high-statistics measurement of the relevant photon fake rates (see for
instance~\cite{Chekanov:2004kz}).

We have also employed our analysis to compare the difference between the fitting functions used by
ATLAS (with the new isolation criteria) given the 2015 and 2016 data sets.  The fitting functions
where extrapolated to invariant masses beyond the TeV region. In summary we have found that:
\begin{itemize}
\item A variation of about 30\% in the extrapolated background near $m_{\gamma\gamma}=1.6\,\TeV$ is
  obtained.
\end{itemize}\par
To conclude, we have extensively examined the status of LHC diphoton searches.  We have compared the
analyses performed on both 2015 and 2016 datasets in order to scrutinize the current state of the
art measurements for systematic effects. Using our approach we have reevaluated the current
experimental sensitivity to beyond standard model physics, especially in the tails of the diphoton
invariant mass distributions, beyond the TeV range. We found that the extrapolation of background
shapes is subject to sizeable uncertainties, potentially affecting the significance of possible
future excesses near the edge of the measured distributions. Furthermore, our analysis motivates
further Monte Carlo studies of the dominant diphoton backgrounds, based on jet flavor tagging
algorithms. Knowledge of whether a jet is of ``quark" or ``gluon" origin would improve our
estimation for the jet-photon faking backgrounds to next to leading order QCD accuracy. It is
important to note that diphoton-based searches at even larger invariant masses, that are highly
motivated, are being performed at present and will continue to be an integral part of the LHC
experimental physics program at the high energy frontier.

\section*{Acknowledgments}

We would like to thank Rikkert Frederix for useful discussions.  J.F.K.~would like to thank CERN for
hospitality while this work was being completed and acknowledges the financial support from the
Slovenian Research Agency (research core funding No. P1-0035).  The work of GP is supported by
grants from the BSF, ISF and ERC and the Weizmann-UK Making Connections Programme. AW is supported
by the DFG cluster of excellence ``Origin and Structure of the Universe'' and the European
Commission (AMVA4NewPhysics, 2020-MSCA-ITN-2015).


\bibliographystyle{JHEP}
\bibliography{TheBib}

\providecommand{\href}[2]{#2}\begingroup\raggedright\begin{thebibliography}{10}

\bibitem{Aaboud:2016tru}
{\scshape {ATLAS}} collaboration, M.~Aaboud et~al., \emph{{Search for
  resonances in diphoton events at $\sqrt{s}$=13 TeV with the ATLAS detector}},
   \href{http://arxiv.org/abs/1606.03833}{{\tt 1606.03833}}.

\bibitem{Khachatryan:2016hje}
{\scshape {CMS}} collaboration, V.~Khachatryan et~al., \emph{{Search for
  resonant production of high-mass photon pairs in proton-proton collisions at
  sqrt(s) = 8 and 13 TeV}},  \href{http://arxiv.org/abs/1606.04093}{{\tt
  1606.04093}}.

\bibitem{ATLAS:2016eeo}
{\scshape ATLAS} collaboration, T.~A. collaboration, \emph{{Search for scalar
  diphoton resonances with 15.4~fb$^{-1}$ of data collected at $\sqrt{s}$=13
  TeV in 2015 and 2016 with the ATLAS detector}}, .

\bibitem{CMS:2016crm}
{\scshape CMS} collaboration, C.~Collaboration, \emph{{Search for resonant
  production of high mass photon pairs using $12.9\,\mathrm{fb^{-1}}$ of
  proton-proton collisions at $\sqrt{s} = 13~\mathrm{TeV}$ and combined
  interpretation of searches at 8 and 13 TeV}}, .

\bibitem{Davis:2016hlw}
J.~H. Davis, M.~Fairbairn, J.~Heal and P.~Tunney, \emph{{The Significance of
  the 750 GeV Fluctuation in the ATLAS Run 2 Diphoton Data}},
  \href{http://arxiv.org/abs/1601.03153}{{\tt 1601.03153}}.

\bibitem{Bondarenko:2016rvd}
K.~Bondarenko, A.~Boyarsky, O.~Ruchayskiy and M.~Shaposhnikov, \emph{{Features
  in the Standard Model diphoton background}},
  \href{http://arxiv.org/abs/1606.0959}{{\tt 1606.0959}}.

\bibitem{Catani:2011qz}
S.~Catani, L.~Cieri, D.~de~Florian, G.~Ferrera and M.~Grazzini, \emph{{Diphoton
  production at hadron colliders: a fully-differential QCD calculation at
  NNLO}}, \href{http://dx.doi.org/10.1103/PhysRevLett.108.072001}{\emph{Phys.
  Rev. Lett.} {\bf 108} (2012) 072001},
  [\href{http://arxiv.org/abs/1110.2375}{{\tt 1110.2375}}].

\bibitem{Campbell:2016yrh}
J.~M. Campbell, R.~K. Ellis, Y.~Li and C.~Williams, \emph{{Predictions for
  diphoton production at the LHC through NNLO in QCD}},
  \href{http://arxiv.org/abs/1603.02663}{{\tt 1603.02663}}.

\bibitem{Aad:2016xcr}
{\scshape { ATLAS}} collaboration, G.~Aad et~al., \emph{{Measurement of the
  inclusive isolated prompt photon cross section in $pp$ collisions at
  $\sqrt{s} = 8$ TeV with the ATLAS detector}},
  \href{http://arxiv.org/abs/1605.03495}{{\tt 1605.03495}}.

\bibitem{Catani:2002ny}
S.~Catani, M.~Fontannaz, J.~P. Guillet and E.~Pilon, \emph{{Cross-section of
  isolated prompt photons in hadron hadron collisions}},
  \href{http://dx.doi.org/10.1088/1126-6708/2002/05/028}{\emph{JHEP} {\bf 05}
  (2002) 028}, [\href{http://arxiv.org/abs/hep-ph/0204023}{{\tt
  hep-ph/0204023}}].

\bibitem{Becher:2013vva}
T.~Becher, G.~Bell, C.~Lorentzen and S.~Marti, \emph{{Transverse-momentum
  spectra of electroweak bosons near threshold at NNLO}},
  \href{http://dx.doi.org/10.1007/JHEP02(2014)004}{\emph{JHEP} {\bf 02} (2014)
  004}, [\href{http://arxiv.org/abs/1309.3245}{{\tt 1309.3245}}].

\bibitem{Becher:2012xr}
T.~Becher, C.~Lorentzen and M.~D. Schwartz, \emph{{Precision Direct Photon and
  W-Boson Spectra at High $p_T$ and Comparison to LHC Data}},
  \href{http://dx.doi.org/10.1103/PhysRevD.86.054026}{\emph{Phys. Rev.} {\bf
  D86} (2012) 054026}, [\href{http://arxiv.org/abs/1206.6115}{{\tt
  1206.6115}}].

\bibitem{Becher:2015yea}
T.~Becher and X.~Garcia~i Tormo, \emph{{Addendum: Electroweak Sudakov effects
  in W, Z and gamma production at large transverse momentum}},
  \href{http://dx.doi.org/10.1103/PhysRevD.92.073011}{\emph{Phys. Rev.} {\bf
  D92} (2015) 073011}, [\href{http://arxiv.org/abs/1509.01961}{{\tt
  1509.01961}}].

\bibitem{Schwartz:2016olw}
M.~D. Schwartz, \emph{{Precision direct photon spectra at high energy and
  comparison to the 8 TeV ATLAS data}},
  \href{http://arxiv.org/abs/1606.02313}{{\tt 1606.02313}}.

\bibitem{Khachatryan:2015ira}
{\scshape {CMS}} collaboration, V.~Khachatryan et~al., \emph{{Comparison of the
  Z/$\gamma$$^{?}$ + jets to $\gamma$ + jets cross sections in pp collisions at
  $ \sqrt{s}=8 $ TeV}}, \href{http://dx.doi.org/10.1007/JHEP04(2016)010,
  10.1007/JHEP10(2015)128}{\emph{JHEP} {\bf 10} (2015) 128},
  [\href{http://arxiv.org/abs/1505.06520}{{\tt 1505.06520}}].

\bibitem{Aad:2013gaa}
{\scshape {ATLAS}} collaboration, G.~Aad et~al., \emph{{Dynamics of
  isolated-photon plus jet production in pp collisions at $\sqrt(s)=7$ TeV with
  the ATLAS detector}},
  \href{http://dx.doi.org/10.1016/j.nuclphysb.2013.07.025}{\emph{Nucl. Phys.}
  {\bf B875} (2013) 483--535}, [\href{http://arxiv.org/abs/1307.6795}{{\tt
  1307.6795}}].

\bibitem{Banfi:2006hf}
A.~Banfi, G.~P. Salam and G.~Zanderighi, \emph{{Infrared safe definition of jet
  flavor}}, \href{http://dx.doi.org/10.1140/epjc/s2006-02552-4}{\emph{Eur.
  Phys. J.} {\bf C47} (2006) 113--124},
  [\href{http://arxiv.org/abs/hep-ph/0601139}{{\tt hep-ph/0601139}}].

\bibitem{Buckley:2015gua}
A.~Buckley and C.~Pollard, \emph{{QCD-aware partonic jet clustering for
  truth-jet flavour labelling}},
  \href{http://dx.doi.org/10.1140/epjc/s10052-016-3925-z}{\emph{Eur. Phys. J.}
  {\bf C76} (2016) 71}, [\href{http://arxiv.org/abs/1507.00508}{{\tt
  1507.00508}}].

\bibitem{Larkoski:2014pca}
A.~J. Larkoski, J.~Thaler and W.~J. Waalewijn, \emph{{Gaining (Mutual)
  Information about Quark/Gluon Discrimination}},
  \href{http://dx.doi.org/10.1007/JHEP11(2014)129}{\emph{JHEP} {\bf 11} (2014)
  129}, [\href{http://arxiv.org/abs/1408.3122}{{\tt 1408.3122}}].

\bibitem{Bauer:2013bza}
C.~W. Bauer and E.~Mereghetti, \emph{{Heavy Quark Fragmenting Jet Functions}},
  \href{http://dx.doi.org/10.1007/JHEP04(2014)051}{\emph{JHEP} {\bf 04} (2014)
  051}, [\href{http://arxiv.org/abs/1312.5605}{{\tt 1312.5605}}].

\bibitem{Agashe:2014kda}
{\scshape Particle Data Group} collaboration, K.~A. Olive et~al., \emph{{Review
  of Particle Physics}},
  \href{http://dx.doi.org/10.1088/1674-1137/38/9/090001}{\emph{Chin. Phys.}
  {\bf C38} (2014) 090001}.

\bibitem{ATLAS:2011kuc}
{ATLAS}, \emph{{Expected photon performance in the ATLAS experiment}},  Tech.
  Rep. ATL-PHYS-PUB-2011-007, CERN, Geneva, Apr, 2011.

\bibitem{Neufeld:2010fj}
R.~B. Neufeld, I.~Vitev and B.~W. Zhang, \emph{{The Physics of
  $Z^0/\gamma^*$-tagged jets at the LHC}},
  \href{http://dx.doi.org/10.1103/PhysRevC.83.034902}{\emph{Phys. Rev.} {\bf
  C83} (2011) 034902}, [\href{http://arxiv.org/abs/1006.2389}{{\tt
  1006.2389}}].

\bibitem{Campbell:1999ah}
J.~M. Campbell and R.~K. Ellis, \emph{{An Update on vector boson pair
  production at hadron colliders}},
  \href{http://dx.doi.org/10.1103/PhysRevD.60.113006}{\emph{Phys. Rev.} {\bf
  D60} (1999) 113006}, [\href{http://arxiv.org/abs/hep-ph/9905386}{{\tt
  hep-ph/9905386}}].

\bibitem{Campbell:2011bn}
J.~M. Campbell, R.~K. Ellis and C.~Williams, \emph{{Vector boson pair
  production at the LHC}},
  \href{http://dx.doi.org/10.1007/JHEP07(2011)018}{\emph{JHEP} {\bf 07} (2011)
  018}, [\href{http://arxiv.org/abs/1105.0020}{{\tt 1105.0020}}].

\bibitem{Campbell:2015qma}
J.~M. Campbell, R.~K. Ellis and W.~T. Giele, \emph{{A Multi-Threaded Version of
  MCFM}}, \href{http://dx.doi.org/10.1140/epjc/s10052-015-3461-2}{\emph{Eur.
  Phys. J.} {\bf C75} (2015) 246}, [\href{http://arxiv.org/abs/1503.06182}{{\tt
  1503.06182}}].

\bibitem{Boughezal:2016wmq}
R.~Boughezal, J.~M. Campbell, R.~K. Ellis, C.~Focke, W.~Giele, X.~Liu et~al.,
  \emph{{Color singlet production at NNLO in MCFM}},
  \href{http://arxiv.org/abs/1605.08011}{{\tt 1605.08011}}.

\bibitem{Alwall:2014hca}
J.~Alwall, R.~Frederix, S.~Frixione, V.~Hirschi, F.~Maltoni, O.~Mattelaer
  et~al., \emph{{The automated computation of tree-level and next-to-leading
  order differential cross sections, and their matching to parton shower
  simulations}}, \href{http://dx.doi.org/10.1007/JHEP07(2014)079}{\emph{JHEP}
  {\bf 07} (2014) 079}, [\href{http://arxiv.org/abs/1405.0301}{{\tt
  1405.0301}}].

\bibitem{Sjostrand:2006za}
T.~Sjostrand, S.~Mrenna and P.~Z. Skands, \emph{{PYTHIA 6.4 Physics and
  Manual}}, \href{http://dx.doi.org/10.1088/1126-6708/2006/05/026}{\emph{JHEP}
  {\bf 05} (2006) 026}, [\href{http://arxiv.org/abs/hep-ph/0603175}{{\tt
  hep-ph/0603175}}].

\bibitem{Cacciari:2011ma}
M.~Cacciari, G.~P. Salam and G.~Soyez, \emph{{FastJet User Manual}},
  \href{http://dx.doi.org/10.1140/epjc/s10052-012-1896-2}{\emph{Eur. Phys. J.}
  {\bf C72} (2012) 1896}, [\href{http://arxiv.org/abs/1111.6097}{{\tt
  1111.6097}}].

\bibitem{Cacciari:2008gp}
M.~Cacciari, G.~P. Salam and G.~Soyez, \emph{{The Anti-k(t) jet clustering
  algorithm}},
  \href{http://dx.doi.org/10.1088/1126-6708/2008/04/063}{\emph{JHEP} {\bf 04}
  (2008) 063}, [\href{http://arxiv.org/abs/0802.1189}{{\tt 0802.1189}}].

\bibitem{Aad:2012tba}
{\scshape {ATLAS}} collaboration, G.~Aad et~al., \emph{{Measurement of
  isolated-photon pair production in $pp$ collisions at $\sqrt{s}=7$ TeV with
  the ATLAS detector}},
  \href{http://dx.doi.org/10.1007/JHEP01(2013)086}{\emph{JHEP} {\bf 01} (2013)
  086}, [\href{http://arxiv.org/abs/1211.1913}{{\tt 1211.1913}}].

\bibitem{Aad:2011mh}
{\scshape {ATLAS}} collaboration, G.~Aad et~al., \emph{{Measurement of the
  isolated di-photon cross-section in $pp$ collisions at $\sqrt{s}=7$ TeV with
  the ATLAS detector}},
  \href{http://dx.doi.org/10.1103/PhysRevD.85.012003}{\emph{Phys. Rev.} {\bf
  D85} (2012) 012003}, [\href{http://arxiv.org/abs/1107.0581}{{\tt
  1107.0581}}].

\bibitem{Chatrchyan:2014fsa}
{\scshape { CMS}} collaboration, S.~Chatrchyan et~al., \emph{{Measurement of
  differential cross sections for the production of a pair of isolated photons
  in pp collisions at $\sqrt{s}=7\,\text {TeV} $}},
  \href{http://dx.doi.org/10.1140/epjc/s10052-014-3129-3}{\emph{Eur. Phys. J.}
  {\bf C74} (2014) 3129}, [\href{http://arxiv.org/abs/1405.7225}{{\tt
  1405.7225}}].

\bibitem{Cowan:2010js}
G.~Cowan, K.~Cranmer, E.~Gross and O.~Vitells, \emph{{Asymptotic formulae for
  likelihood-based tests of new physics}},
  \href{http://dx.doi.org/10.1140/epjc/s10052-011-1554-0,
  10.1140/epjc/s10052-013-2501-z}{\emph{Eur. Phys. J.} {\bf C71} (2011) 1554},
  [\href{http://arxiv.org/abs/1007.1727}{{\tt 1007.1727}}].

\bibitem{Chekanov:2004kz}
{\scshape {ZEUS}} collaboration, S.~Chekanov et~al., \emph{{Substructure
  dependence of jet cross sections at HERA and determination of alpha(s)}},
  \href{http://dx.doi.org/10.1016/j.nuclphysb.2004.08.049}{\emph{Nucl. Phys.}
  {\bf B700} (2004) 3--50}, [\href{http://arxiv.org/abs/hep-ex/0405065}{{\tt
  hep-ex/0405065}}].

\end{thebibliography}\endgroup

\end{document}
